\documentclass[reprint,floatfix,superscriptaddress]{revtex4-1}
\usepackage{fullpage}
\usepackage{amssymb}
\usepackage{graphicx}
\usepackage{graphics}
\usepackage{subfigure}
\usepackage{dsfont}
\usepackage{mathrsfs}
\usepackage{amsmath}
\usepackage{latexsym}
\usepackage{amsfonts}
\usepackage{amsthm}

\begin{document}
\title{Accelerating cycle expansions by dynamical conjugacy}
\author{ Ang Gao }
\email[E-mail: ]{ang.dionysos.gao@gmail.com}
\affiliation{The
Department of Physics, Tsinghua University}
\author{Jianbo Xie}
\affiliation{The Department of Physics, Tsinghua University}
\affiliation{The Department of Physics, UC Berkeley}
\author{Yueheng Lan}
\email[E-mail: ]{lanyh@mail.tsinghua.edu.cn}
\affiliation{The
Department of Physics, Tsinghua University}

\begin{abstract}

Periodic orbit theory provides two important functions---the
dynamical zeta function and the spectral determinant for the
calculation of dynamical averages in a nonlinear system. Their cycle
expansions converge rapidly when the system is uniformly hyperbolic
but greatly slowed down in the presence of non-hyperbolicity. We
find that the slow convergence can be associated with singularities
in the natural measure. A properly designed coordinate
transformation may remove these singularities and results in a
dynamically conjugate system where fast convergence is restored. The
technique is successfully demonstrated on several examples of
one-dimensional maps and some remaining challenges are discussed.
\end{abstract}
\maketitle
\section{Introduction}
Equilibrium statistical physics has been extremely successful, while
accurate computation of physical averages in non-equilibrium systems
remains a great challenge both theoretically and practically, due to
the intrinsic difficulty of procuring the right statistical weight
of various states based on equations of motion that govern system
evolution~\cite{cvt89b,hao90b}. From a dynamical systems point of
view, the statistical weight is proportional to the natural measure
of states in the phase space, which often is non-smooth or even
singular in a chaotic system~\cite{sinai76,frisch} and thus flunks
an accurate representation. Fortunately, periodic orbit theory (POT)
avoids a direct description of the possibly fractal measure by
expressing phase space averages in terms of averages on periodic
orbits or cycles and thus is a powerful way for reliable and
accurate characterization~\cite{DasBuch,gutbook,so96per,mic97un} of
a nonlinear chaotic system. The associated cycle expansion of
spectral determinants or dynamical zeta functions orders cycles in a
hierarchical way such that dynamical averages are dominated by a few
short cycles and the longer ones give decreasing corrections. For a
uniformly hyperbolic system with finite symbolic dynamics, the
corrections decrease exponentially or even
super-exponentially~\cite{cexp,hhrugh92,hat95rat}. However, for
non-hyperbolic systems, the convergence could be extremely poor,
which severely limits the application of cycle
expansions~\cite{DasBuch}.

Real physical systems are non-hyperbolic in different cases. For
example, in an intermittent system, typical trajectories alternate
between regular and chaotic motions in an irregular way and thus
cause non-hyperbolicity~\cite{carl97int,art03int}. A milder type of
non-hyperbolicity is created by the strong contraction at specific
points such as homoclinic tangencies in the H\'enon map or critical
points in 1-d maps~\cite{cexp,rob90rec}. As a consesuence, the
natural measure becomes non-analytic and thus the nice shadowing
property among cycles that is necessary for fast convergence of
cycle expansions fades out. Poles appear near the origin of the
complex plane in the dynamical zeta function and the spectral
determinant, which gives much trouble to a polynomial approximation
as in the normal cycle expansion. To compute averages with fair
accuracy, many cycles are needed, which usually requires
unaffordable amount of resources. Thus, how to accelerate the
convergence of cycle expansions in the presence of non-hyperbolicity
is a key problem in practice.

Several accelerating schemes have been proposed based on the
analyticity of the spectral functions. One idea is to identify and
remove the poles that are near the origin and thus expand the radius
of convergence. In \cite{cexp,rob90rec,aur90con}, the dominant terms
in the tail of the expansion are estimated and summed up to
approximately determine the leading pole. More accurate estimation
is obtained by using Pad{\' e} approximation, which is valid not
only for  computing the leading pole but also for seeking other
ones~\cite{eck93res,main99sem}. An interesting consequence of
analyticity of the underlying dynamics is the existence of infinite
sum rules which symptom strong correlations among periodic orbits.
These exact relations show signs of information redundancy embedded
in the whole set of periodic orbits and can be utilized to
accelerate the convergence of cycle expansions~\cite{sun99per}. In
certain cases, analyticity can be used to derive the spectral
function with no resort to periodic orbits and thus implies a
potential alternative route to the spectrum computation. But so far,
it succeeded only for several very specific maps and hard to be
generalized to other examples~\cite{cv98bey}.

In this paper, we employ a geometric picture of the cycle expansion
to treat the convergence problem for 1-d maps~\cite{skeleton,inv}.
Maps with critical points have a natural measure with singularities,
due to the strong contraction around critical points. This
contraction also deteriorates the dynamical shadowing between cycles
of different lengths and thus leads to a slow convergence of cycle
expansions. So, the singularity in the natural measure is an
effective indicator of unbalance of cycle weights and signals a
small radius of convergence. One idea for expediting convergence is
thus to identify and then clear out singularities in the natural
measure. In the current paper, we achieve this by properly designing
coordinate transformations such that the resulted conjugate map has
a natural measure with no singularity. The computation of dynamical
averages in the original map can be efficiently done with
counterparts in the conjugate map since the convergence is much
improved in the new map.

In the following, after a brief review of periodic orbit theory in
Section~\ref{sec:pot}, we discuss in Section~\ref{sec:Icce} the
convergence of the dynamical zeta function and the spectral
determinant for maps with critical points. A comparison between the
two spectral functions is made and the importance of hyperbolicity
for efficient calculation is emphasized. With a description of the
geometric significance of the truncation in cycle expansions, our
accelerating scheme is presented and tested on several examples. In
Section~\ref{sec:sum}, we summarize the paper and discuss the
existing problems  and possible directions for further
investigation.

\section{Periodic orbit theory \label{sec:pot}}
More often than not, dynamical averages are conveniently computed
via time averaging,
\begin{equation}
\overline{a(x_0)}=\frac{1}{N} \sum_{i=0}^{N} a(x_i)\,,\label{eq:ta}
\end{equation}
where $x_0 \to x_1 \to \cdots x_i \to \cdots x_N$ is an itinerary
generated by the map $f(x)$ and $N$ is a large number. Time
averaging is easy to do but hard to achieve high accuracy. In the
presence of non-hyberbolicity, its convergence is very slow and the
result becomes unreliable. To better understand the dynamics for
more efficient calculation, the phase space average $\langle a
\rangle$ is introduced and has the nice property $\langle a \rangle
= \overline{a(x_0)}$ in an ergodic system. Thus, a geometric picture
is enabled to explore the averaging process, which will be explained
in more detail.

For an ergodic map $f(x)$, when $n \to \infty$, under the map action
any smooth initial measure will approach an invariant measure,
called the natural measure.  Formally,
\begin{equation}
\rho(x) = \lim_{n\to \infty} \int_{\mathcal{M}} dy \delta(x-f^n(y))
\rho_0(y) \,,
\end{equation}
where $\rho(x)$ is the natural measure and $\rho_0(y)$ is an initial
smooth measure. With the natural measure, the average $\langle a
\rangle $ can be obtained by
\begin{equation}
\langle a \rangle = \int_{\mathcal{M}} a(x) \rho(x) dx \,,
\end{equation}
which is the most common way in statistical physics for computing
averages. In a chaotic system, however, the measure $\rho(x)$, being
often singular and supported on a fractal set, is hard to obtain,
which motivates introduction of periodic orbit theory.

For a map $f: \mathcal{M} \to \mathcal{M}$ and an observable $a(x)$,
we define the evolution operator $\mathcal{L}^n$: $\mathcal{L}^n
\circ g(y) = \int_{\Omega} \mathcal{L}^n(y,x) g(x) dx$ for any
function $g(y)$. The kernel is
\begin{equation}
\mathcal{L}^n(y,x)=\delta(y-f^n(x))e^{\beta A^n}\,,
\end{equation}
where $\beta$ is an auxiliary variable and $A^n =
\sum_{k=0}^{n-1}a(f^k(x))$. The average $\langle a \rangle $ or
other dynamical properties can be conveniently obtained by virtue of
the spectrum of $\mathcal{L}$. Suppose the leading eigenvalue of
$\mathcal{L}$ is $e^{s_0}$, then we have~\cite{DasBuch}
\begin{equation}
\langle a \rangle = \frac{\partial s_0}{\partial \beta}\mid_{\beta=0}\,. \label{eq:aver}
\end{equation}
Specifically, when $\beta=0$, we have
$\mathcal{L}^n(y,x)=\delta(y-f^n(x))$, which is the kernel of the
so-called Perron-Frobenius operator. The escape rate of a dynamical
system, which we denote by $\gamma$, can be obtained by computing
the leading eigenvalue of this operator
\begin{equation}
\gamma = - s_0\,.
\end{equation}

The eigenvalues of the evolution operator $\mathcal{L}$ can be
detected with the help of the spectral determinant, which is related
to the trace of  $ \mathcal{L}$ and thus to the periodic orbits by
the identity
\begin{equation}
\ln \det(1-z\mathcal{L}) = \mathrm{Tr} \ln(1-z\mathcal{L})\,.
\end{equation}
For one-dimensional maps, detailed manipulation shows that~\cite{DasBuch}
\begin{equation}
\det(1-z\mathcal{L}) = \exp(-\sum_p \sum_{r=1}^{\infty}\frac{1}{r}\frac{z^{n_pr}e^{r\beta A_p}}{|1-\Lambda_p^r|})\,,
\end{equation}
where $ p$ denotes prime cycles which are not repeats of  shorter
ones. $\Lambda_p$ is the stability eigenvalue of cycle $p$ and $n_p$
is its length. In most classical computations, we are only
interested in the leading eigenvalue. Obviously, the smallest
positive zero of the above-defined spectral determinant is
$e^{-s_0}$, the inverse of the leading eigenvalue of $\mathcal{L}$.
In view of Eq.~(\ref{eq:aver}), we are able to compute  dynamical
averages with the spectral determinant.

The leading eigenvalue can alternatively be obtained from a simpler spectral function---the dynamical zeta function
\begin{equation}
\frac{1}{\zeta} = \prod_p(1-t_p)\,,t_p= \frac{1}{|\Lambda_p|}z^{n_p}e^{\beta A_p}\,.
\end{equation}
It can be proved that $\frac{1}{\zeta}$ is the zeroth-order
approximation of the spectral determinant and has kept the smallest
positive zero unchanged. Most often, however, they have different
analytic properties.

Practically, to evaluate zeros, we expand the spectral determinant
or the dynamical zeta function in terms of power series and get
polynomial approximation through truncation. The power series
expansion is one type of cycle expansion. For example, for the
one-dimensional map with complete binary symbolic dynamics, the
dynamical zeta function can be expanded as
\begin{equation}
\begin{array}{lll}
\frac{1}{\zeta} & = & 1-t_0-t_1-[(t_{01}-t_0t_1)]\\
& & -[(t_{001} - t_{01}t_0)+(t_{011}-t_{01}t_1)] -\cdots\,.
\end{array}
\label{eq:ce}
\end{equation}

If we keep only the terms explicitly shown in Eq.~(\ref{eq:ce}), its
cycle expansion is truncated at cycle length $3$ and results in a
polynomial in $z$ of degree $3$. The linear term is the fundamental
contribution which gives the dominant part of the expansion. Higher
order terms are curvature corrections which consist of contributions
from prime cycles such as $t_{01}$ and from pseudo-cycles being
combination of prime cycles such as $t_0t_1$. The cancelation
between cycles and pseudo-cycles signals shadowing properties and
smoothness of the underlying dynamics and results in an exponential
decrease of the curvature corrections when uniform hyperbolicity is
assumed. However, there is no good cancelation in the presence of
non-hyperbolicity and as a result cycle expansion converges very
slowly. In the current paper, we are trying to restore this
cancelation by dynamical conjugacy under certain circumstances.

\section{Improving the convergence of cycle expansions \label{sec:Icce}}

\subsection{Notes on numerical computation}
To calculate dynamical averages, we need to build the truncated
version of the  dynamical zeta function and the spectral
determinant. The detailed explanation for an efficient computation
can be found in~\cite{DasBuch}, which is omitted here for brevity.
With a truncation length $N$, we drop out all the cycles longer than
$N$. To study the convergence of cycle expansions, as an example, we
will evaluate the escape rate and other dynamical averages with the
truncated dynamical zeta function and spectral determinant. Also, we
will check how the computational error of physical averages depends
on the truncation length. However, except for very few cases, we
cannot obtain the exact average values. Therefore, we use averages
obtained with the truncation length $N_{\max}+1$ as the ``exact''
values when estimating errors with the truncation length no larger
than $N_{\max}$. Figures are plotted to show this dependence and the
logarithmic scale is often used in the ordinate.

The natural measure is computed by map iterations. We choose a
random initial point and iterate it many times, usually $10^7$ if
not specified otherwise. Then, by counting the times that the point
enters a small interval, we get a probability distribution, which is
a numerical approximation of the natural measure for an ergodic
system. As the maps discussed in this paper are all ergodic, though
not very accurate in some cases, this method is simple for getting a
rough picture of the natural measure. In addition, with
Eq.~(\ref{eq:ta}),  physical averages are easily computed with the
iterations at the same time.

\subsection{Comparison of the dynamical zeta function and the spectral determinant}
The escape rate and other dynamical averages may be evaluated with
the dynamical zeta function or the spectral determinant. However,
the  convergence rates of the two methods are quite different, which
is due to the difference in their radius of convergence in cycle
expansions.  To show this, we  calculate the escape rate of the map
$ f(x) = 6x(1-x),\, \mathcal{M}=[0,1],\,f: \mathcal{M} \to
\mathcal{M}$ with the dynamical zeta function and the spectral
determinant. The results are listed in TABLE~\ref{ta:2}.
\begingroup
\squeezetable
\begin{table}[htp]
\begin{tabular}{|l|l|l|}
\hline
N & $ \gamma(\frac{1}{\zeta})$ & $ \gamma(\det(1-z\mathcal{L}))$ \\
\hline
1 &0.87& 0.9\\
\hline
2 & 0.83 & 0.83 \\
\hline
3 & 0.83151 & 0.831492\\
\hline
4 & 0.831492 & 0.831492987\\
\hline
5 & 0.831493012 & 0.831492987487621\\
\hline
6 & 0.831492987 & 0.831492987487621617307\\
\hline
7 & 0.8314929875 & 0.8314929874876216173072762950\\
\hline
8 & 0.831492987487 & 0.83149298748762161730727629503691\\
\hline
9 & 0.8314929874876 & \\
\hline
10 & 0.83149298748762 & \\
\hline
\end{tabular}
\caption{Escape rate obtained by the dynamical zeta function $\frac{1}{\zeta}$ and the spectral determinant $\det(1-z\mathcal{L})$ for the map $f(x)=6x(1-x)$ on the interval $[0,1]$.}
\label{ta:2}
\end{table}
\endgroup
According to the table, both methods converge fast, thanks to the
nearly perfect cancelation between prime and pseudo-cycles. Thus
very accurate results can be obtained with only several short prime
cycles. Moreover,  the results computed with the spectral
determinant converge much more quickly than with the dynamical zeta
function, implying a difference in their analyticity.
\begin{figure}[htp]
\subfigure[the error of the escape rate computed with the dynamical zeta function]{\includegraphics[width=0.48\textwidth,height=0.3\textwidth]{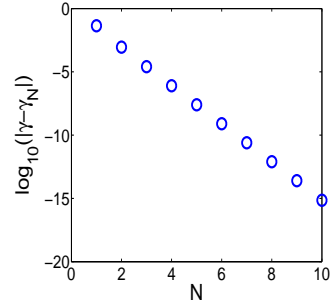}}
\subfigure[the error of the escape rate computed with the spectral determinant]{\includegraphics[width=0.48\textwidth,height=0.3\textwidth]{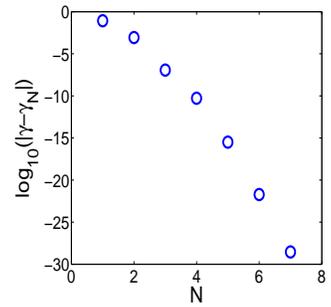}}
\caption{The error of the escape rate for the map $f(x)=6x(1-x)$ computed with (a) the dynamical zeta function and (b) the spectral determinant.}
\label{fig:1}
\end{figure}

FIG.~\ref{fig:1} shows the dependence of the error in the computed
escape rate on the truncation length $N$  with the dynamical zeta
function and the spectral determinant.  It is clear that the
logarithm of the error decreases linearly for the dynamical zeta
function and super-linearly for the spectral determinant, which
suggests an exponential and a super-exponential decrease in the
error itself, respectively. The reason for this difference is that
the spectral determinant is analytic over the whole complex plane,
while the dynamical zeta function is only analytic in a region with
a finite radius. Thus, the coefficient of the $N$th-order term
decreases super-exponentially with $N$ in the spectral determinant,
and exponentially in the dynamical zeta function. From a pure
algebraic point of view, the way in which the coefficient of the
$N$th-order term decreases determines the convergence rate.

\subsection{The influence of hyperbolicity }

Having compared the convergence rate of  the dynamical zeta function
and the spectral determinant for the map $ f(x) = Ax(1-x)$ with
$A=6$, we check how the value of $A$  influences the  convergence
rate.

\begin{figure}[htp]
\subfigure[the error of the escape rate computed with the dynamical zeta function]{\includegraphics[width=0.48\textwidth,height=0.3\textwidth]{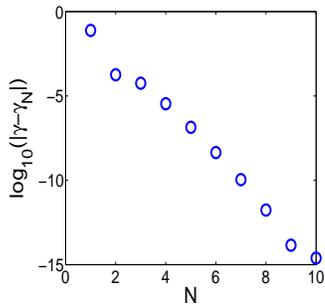}}
\subfigure[the error of the escape rate computed with the spectral determinant]{\includegraphics[width=0.48\textwidth,height=0.3\textwidth]{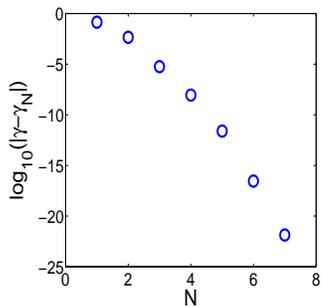}}
\caption{The error of the escape rate for the map $f(x)=5x(1-x)$ computed with (a) the dynamical zeta function and (b) the spectral determinant.}
\label{fig:2}
\end{figure}

\begin{figure}[htp]
\subfigure[the error of the escape rate computed with the dynamical zeta function ]{\includegraphics[width=0.48\textwidth,height=0.3\textwidth]{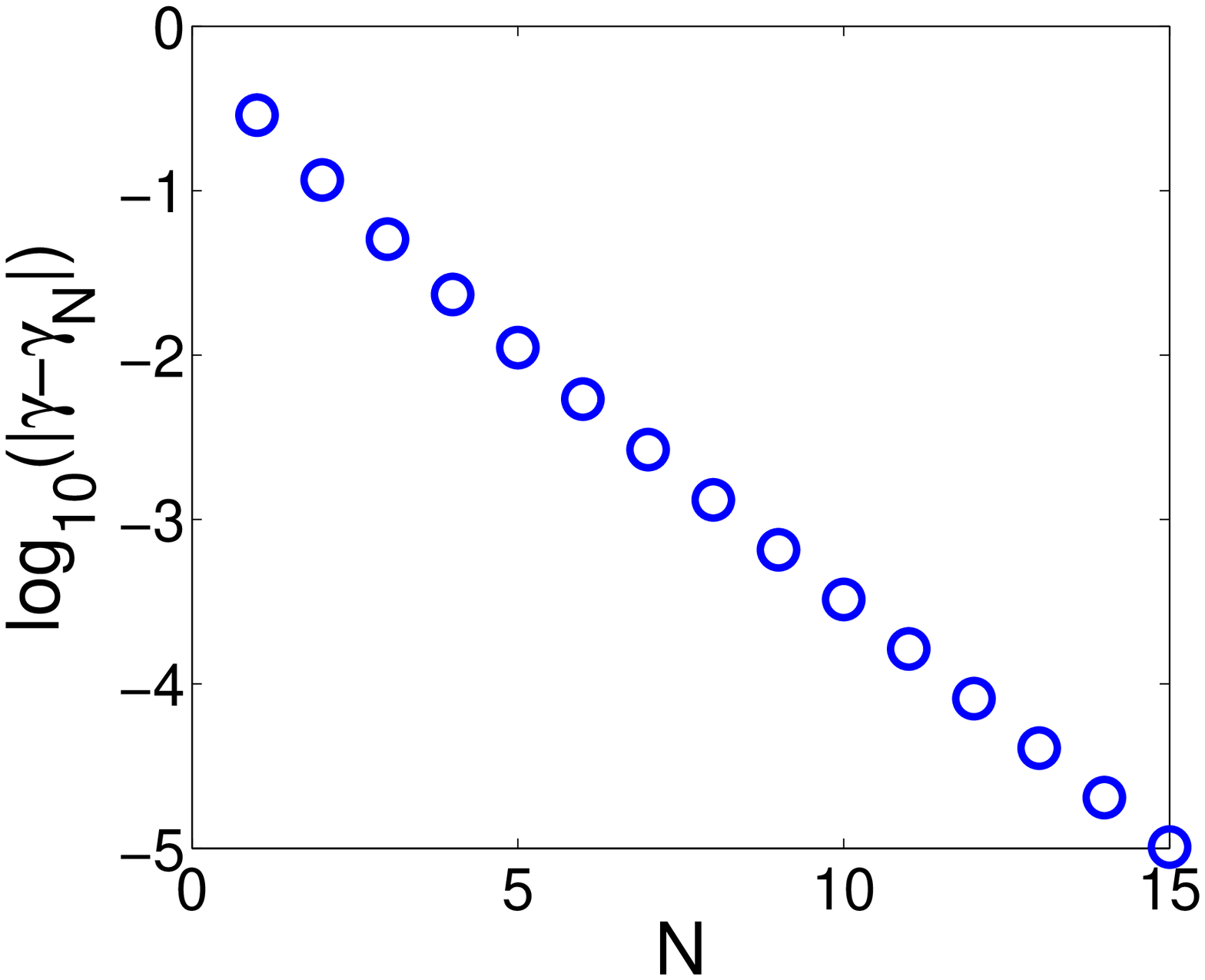}} \quad
\subfigure[the error of the escape rate computed with the spectral determinant]{\includegraphics[width=0.48\textwidth,height=0.3\textwidth]{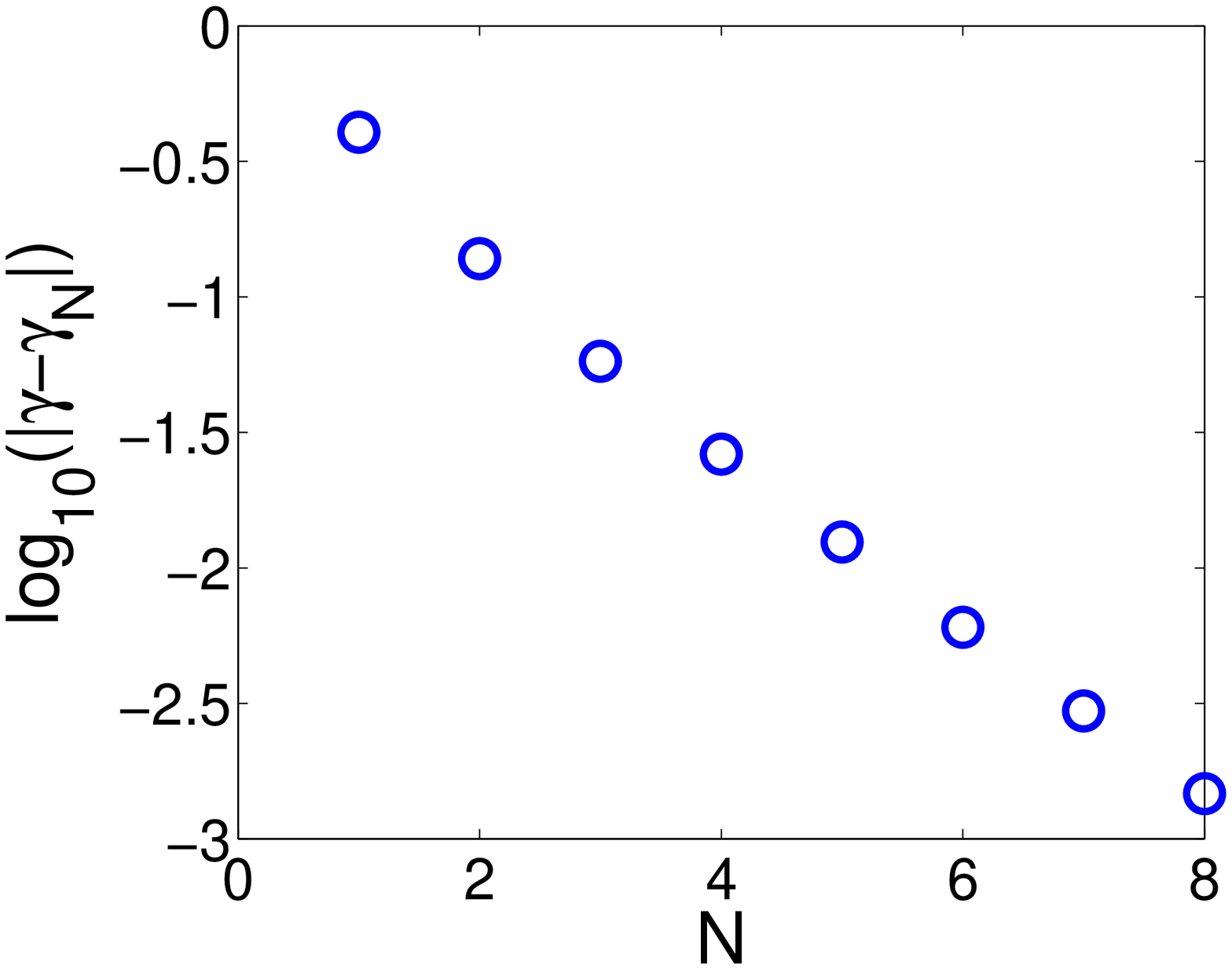}}
\caption{The error of the escape rate for the map $f(x)=4x(1-x)$ computed with (a) the dynamical zeta function and (b) the spectral determinant.}
\label{fig:3}
\end{figure}

First, we set $A$ equal to $5$ and repeat the above computation for
the escape rate, the results are shown in FIG.~\ref{fig:2}. We see
that though a little slower than the $A=6$ case, both methods
converge  fast---the dynamical zeta function method converges nearly
exponentially and  the spectral determinant exhibits a beautiful
super-exponential convergence, just like what happened before.  It
looks as if  the change of $A$ had little effect on convergence.
However, if we set $A$ equal to $4$, a dramatic change happens to
the convergence. As shown in FIG.~\ref{fig:3}, the dynamical zeta
function  with cycles up to length $15$ gives  an error of
$10^{-5}$, while in the $A=5$ case, the error is $ 10^{-15}$.
Moreover, the results obtained by the spectral determinant even lose
the super-exponential convergence and exhibit only an exponential
convergence. This phenomena implies that, in this special case, the
spectral determinant may not be an entire function any more.

\begin{figure}[htp]
\subfigure[the error of the escape rate computed with the dynamical zeta function]{\includegraphics[width=0.48\textwidth,height=0.3\textwidth]{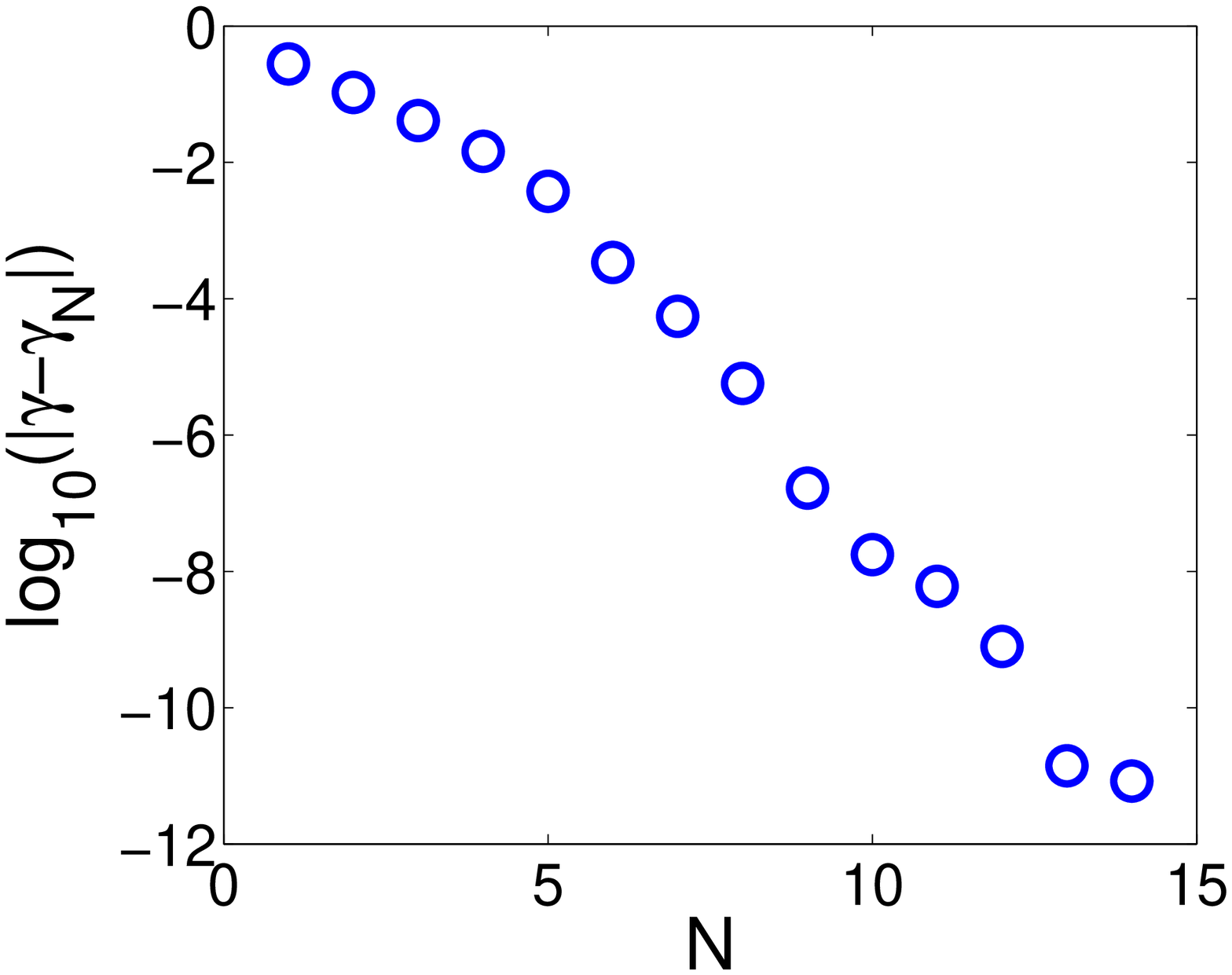}} \quad
\subfigure[the error of the escape rate computed with the spectral determinant]{\includegraphics[width=0.48\textwidth,height=0.3\textwidth]{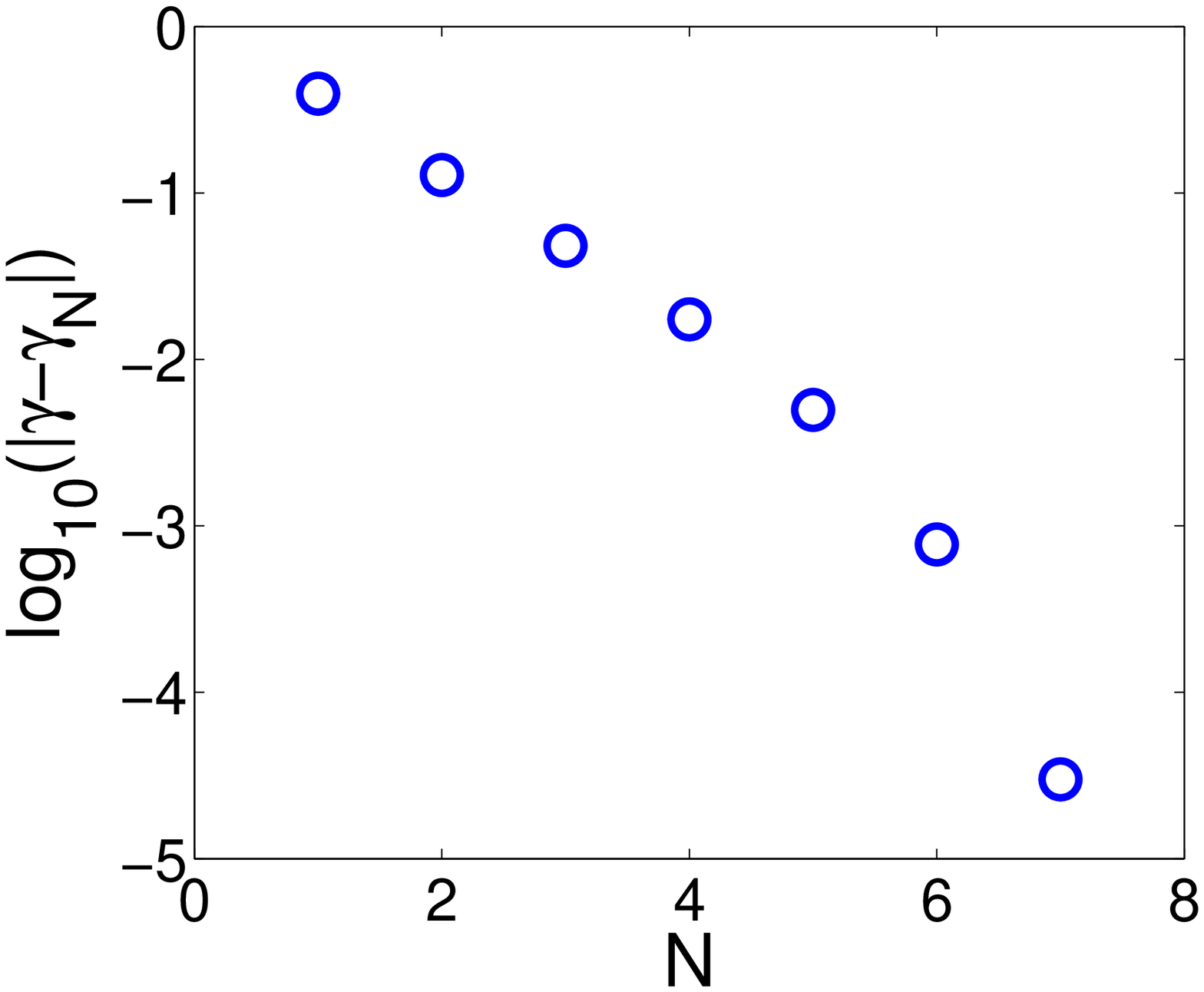}}
\caption{The error of the escape rate  for the map $f(x)=4.001x(1-x)$ computed with (a) the dynamical zeta function and (b) the spectral determinant.}
\label{fig:4}
\end{figure}
If we increase $A$ from $4$  to $4.001$, the convergence of the
expansion improves dramatically, as depicted in Figure \ref{fig:4}.
When $A=4.001$, the super-exponential convergence of the spectral
determinant is restored. An apparent property that makes the map of
$ A=4$ different is that the height of the critical point falls
within the interval $[0,1]$, which is known to be the cause of the
slow convergence~\cite{DasBuch}. Here, we study this phenomena in
great detail and will design a technique to counter its effect
later. To show how the critical point influences the convergence of
cycle expansions, we use the dynamical zeta function to calculate
the escape rate for maps with a higher-order critical point. One
general form of such maps is
$f(x)=1-|2x-1|^k,\,x\in[0,1]$~\cite{det09sto}.

\begin{figure}[htp]
\includegraphics[width=0.48\textwidth,height=0.3\textwidth]{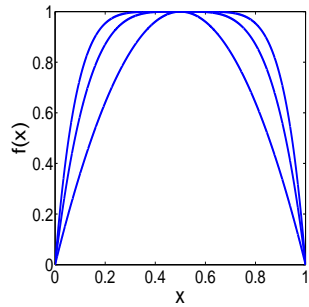}
\caption{The graph of $f(x)=1-(2x-1)^k,\,k=2,4,6$.}
\label{fig:5}
\end{figure}

\begin{figure}[htp]
\subfigure[the error of the escape rate for the map $f(x)=1-(2x-1)^4$]{\includegraphics[width=0.48\textwidth,height=0.3\textwidth]{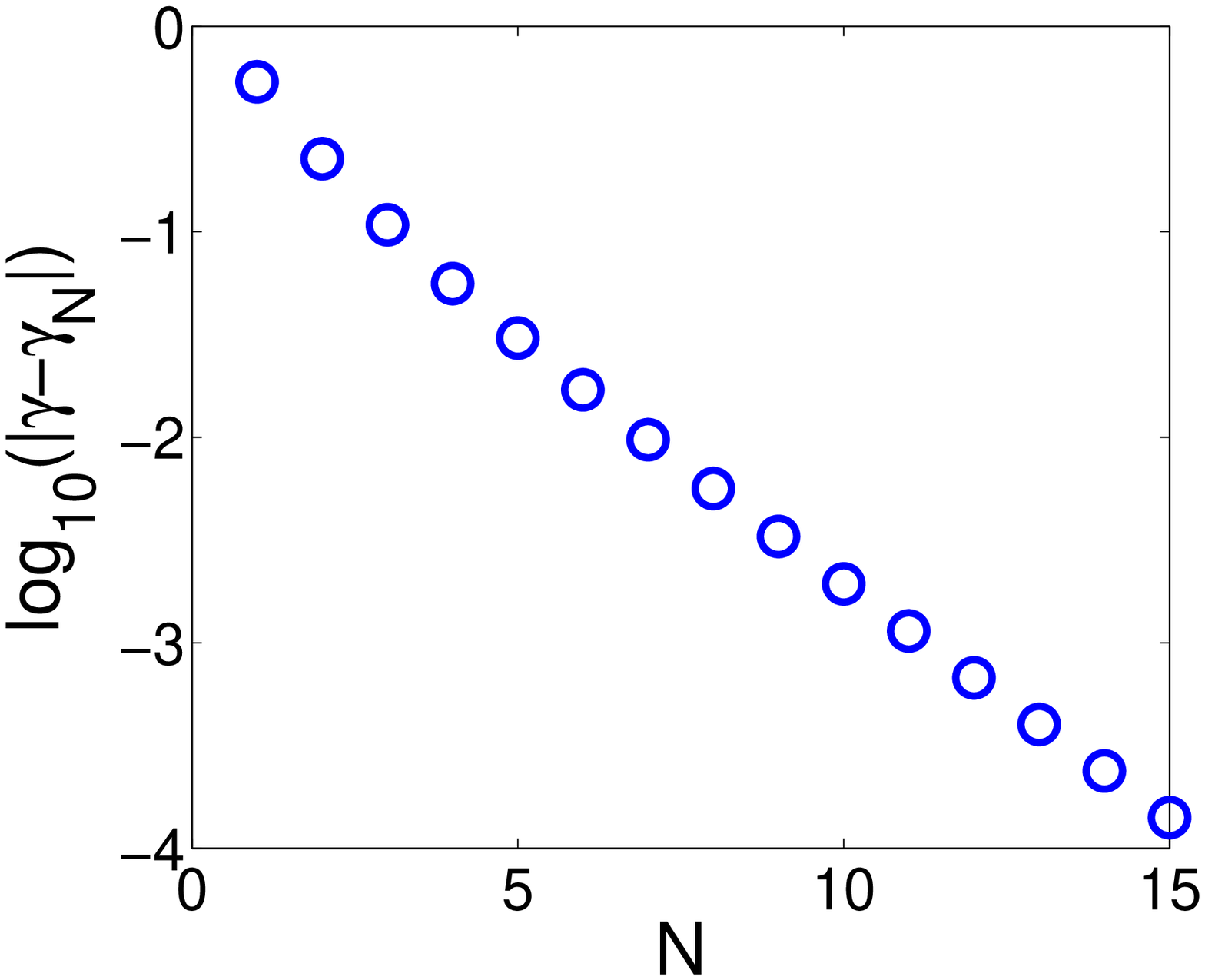}} \quad
\subfigure[the error of the escape rate for the map $f(x)=1-(2x-1)^6$]{\includegraphics[width=0.48\textwidth,height=0.3\textwidth]{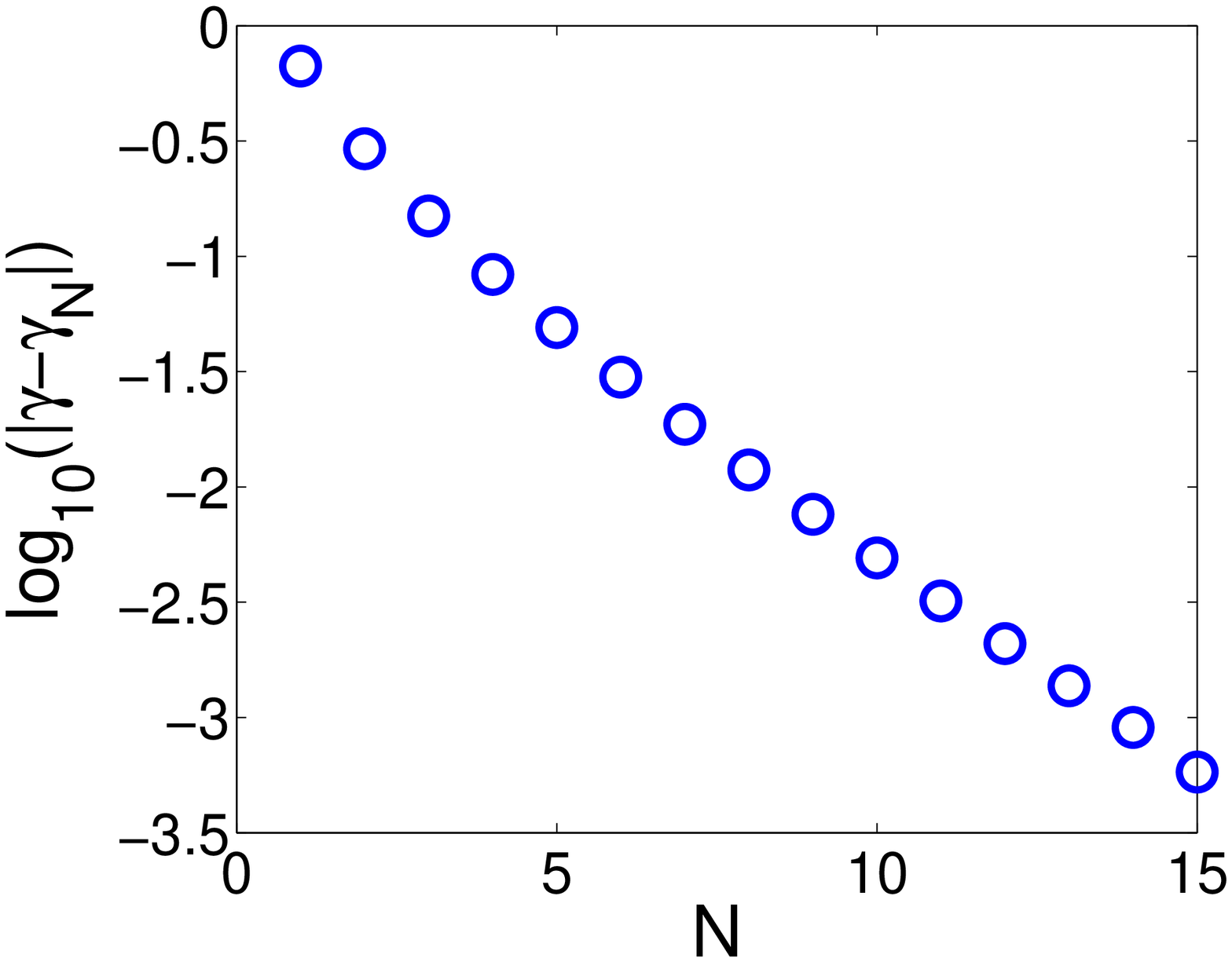}}
\caption{The error of the escape rate for maps with a higher-order critical point computed with the dynamical zeta function.}
\label{fig:6}
\end{figure}

In FIG.~\ref{fig:5},  the profile of the map for different $k$ is portrayed. As the order $k$ increases,
the top of the map profile becomes flatter and flatter. The error of the
escape rate obtained from the dynamical zeta function is displayed
in FIG.~\ref{fig:6} and FIG.~\ref{fig:3}(a), where we can see that
the convergence is  poorer for  maps with a higher-order critical
point.  To know why the dynamical zeta function and the spectral
determinant flaw for maps with critical points, we must have a
clear understanding of the nature of the approximations when we
apply a truncation to the dynamical zeta function.

\subsection{The significance of the truncated dynamical zeta function}
\begin{figure}[htp]
\includegraphics[width=0.48\textwidth,height=0.3\textwidth]{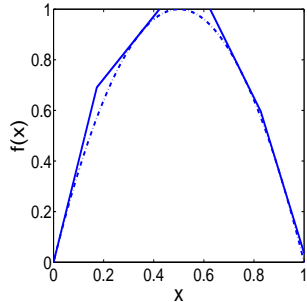}
\caption{A piecewise linear map approximation of the map $f(x)=4x(1-x)$.}
\label{fig:7}
\end{figure}

As the number of all prime cycles is infinite, the truncation to the
spectral functions is needed for an efficient computation. For
example, for the unimodal maps discussed above, if we truncate at
the shortest cycles, the dynamical zeta function is $
\frac{1}{\zeta} = 1 - t_0- t_1 $. This truncated dynamical zeta
function leaves out all the curvature corrections and is a rough
approximation for the original map. Now, one question can be asked:
is there a linear map having $\frac{1}{\zeta}= 1 - t_0 -t_1 $ as its
truncated first-order dynamical zeta function? A simple example is a
piecewise linear map consisting of two branches, which is
constructed in this way: firstly, find the fixed points of the map
along with the slopes at these points, then, draw line segments
which pass the fixed points and are tangent to the graph of the map.
If we want to construct a piecewise linear map which has a dynamical
zeta function identical with that of the original map up to order
$N$, we need the positions and slopes of all the periodic points of
period  not larger than $N$. FIG.~\ref{fig:7} displays such a
piecewise linear map for the truncation up to length two, while the
original map is $f(x)=4x(1-x)$.

For the piecewise linear map, the curvature corrections with order
lager than $N$  are nearly zero due to its linearity, and therefore,
we regard it as a prototyped geometric model of the $N$th-order
truncated dynamical zeta function. The  dynamical averages computed
with the $N$th-order truncated dynamical zeta function are very
close to those given by the piecewise linear map. So, how well the
piecewise linear map approximates the original map determines the
computational accuracy of the truncation.

\begin{figure}[htp]
\includegraphics[width=0.48\textwidth,height=0.3\textwidth]{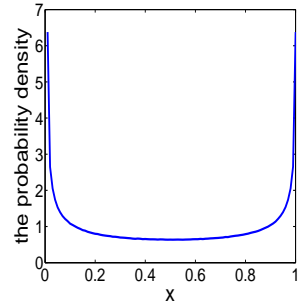}
\caption{The natural measure of the map $f(x)=4x(1-x)$.}
\label{fig:8}
\end{figure}

We know that the average obtained from the dynamical zeta function
is the phase space average $\langle a \rangle = \int_\mathcal{M} dx
\rho(x) a(x)$, where $\rho(x)$ is the natural measure. For example,
for the tent map, the natural measure is uniform everywhere.
However, for the map with critical points, the natural measure has
singularities as portrayed in FIG.~\ref{fig:8}.

It's just these singularities that lead to the slow convergence in a
cycle expansion, because the piecewise linear map obtained from the
truncated dynamical zeta function can not capture well the natural
measure with singularities. Though the natural measure of the
piecewise linear map gets closer and closer to that  of the original
map with increasing truncation length, it fails at the singularity.
As a result, the average obtained from the piecewise linear map, or
equivalently, from the truncated dynamical zeta function for a
non-hyperbolic system, does not converge as fast as in the uniformly
hyperbolic case.

\subsection{Accelerating convergence in the presence of a critical point}
The singularity in the natural measure causes the slow convergence of the dynamical zeta function.  A natural cure of  this trouble is to clear  out the singularity by a coordinate transformation, which results in a new map conjugate to the original one but without a singularity in the natural measure. Consequently, we are able to accelerate the convergence of the cycle expansion with the help of the conjugate map.

\subsubsection{Clear out the singularities \label{sec:cos}}
For maps with a critical point, such as $f(x)= 1-|2x-1|^k,\,k=2,4,6,\cdots$, the probability distribution has an algebraic form in the neighborhood of the singularity, more explicitly
\begin{equation}
\rho(x) \sim \frac{1}{x^{\frac{k-1}{k}}}\,\,\textrm{near $x=0$}\,,
\end{equation}
where $ k$ is the order of the critical point. For example, the natural measure of the map $f(x)=1-(2x-1)^2$ has two singular points: $0$ and $1$, with the probability distribution  $\rho(x) \sim \frac{1}{\sqrt{x}}$ near $ x=0$ and $ \rho(x) \sim \frac{1}{\sqrt{1-x}}$ near $ x=1$.

 A coordinate transformation is needed which stretches the coordinate axis around the singularity in order to remove it. We are able to construct a  homeomorphism $ h:\mathcal{M} \to \mathcal{M}$ of the form $h(x)\propto{|x-x_0|}^{1/k}$ in the neighborhood of the singularity $ x=x_0$ to achieve this goal. For the map $f(x)=1-(2x-1)^2=4x(1-x),\,[0,1] \to [0,1]$, an appropriate transformation is $h(x)=\frac{2}{\pi} \arcsin{\sqrt{x}}$, which as you can see has the desired form  in the neighborhood of $0$ and $1$.

\begin{figure}[htp]
\centering
\subfigure[ $f(x)= 4x(1-x)$]{\includegraphics[width=0.48\textwidth,height=0.3\textwidth]{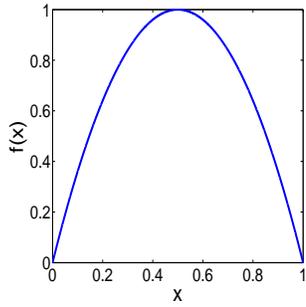}}\quad
\subfigure[the conjugate map]{\includegraphics[width=0.48\textwidth,height=0.3\textwidth]{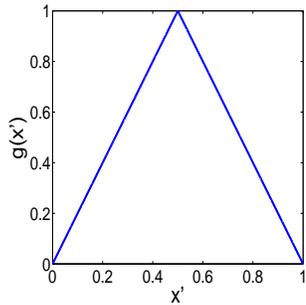}}
\caption{ (a)The map $f(x)=4x(1-x)$ and (b) its conjugate map.}
\label{fig:9}
\end{figure}

With the coordinate transformation, the original map is changed to its conjugate. Denoting the original map by $f(x)$ and the conjugate by $g(x')$, we have the relationship that $f=h^{-1} \circ g \circ h$, or equivalently, $g=h \circ f \circ h^{-1}$. The map $f(x)=1-(2x-1)^2=4x(1-x)$ and its conjugate are displayed in FIG.~\ref{fig:9}.

\begin{figure}[htp]
\centering
\subfigure[the natural measure of map $f(x)= 4x(1-x) $]{\includegraphics[width=0.48\textwidth,height=0.3\textwidth]{spd-4xl1-xr.eps}}\quad
\subfigure[the natural measure of the conjugate map]{\includegraphics[width=0.48\textwidth,height=0.3\textwidth]{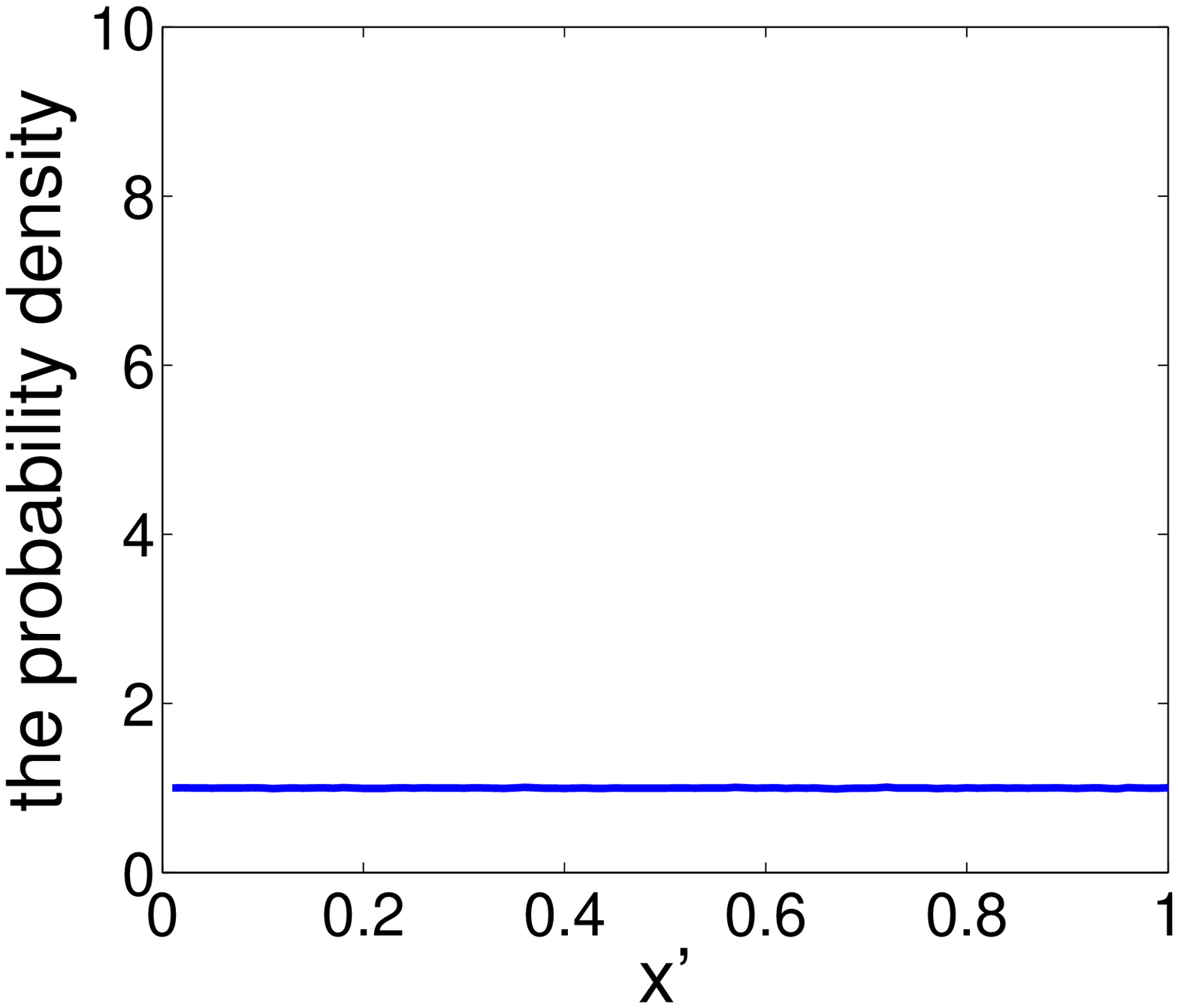}}
\caption{The natural measure (a) of the map $f(x)=4x(1-x) $ and (b) of its conjugate map.}
\label{fig:10}
\end{figure}

The conjugate map has no critical point any more, and the natural measure produced by the two maps are displayed in FIG.~\ref{fig:10}, in which we can see that the natural measure of the conjugate map has no singularity any more.

\subsubsection{The conjugate dynamical zeta function}
 In an ergodic system,  the dynamical average $\langle a \rangle$ can be obtained through the time averaging, as well as an  average on the natural measure as discussed before.

  An ergodic map $f:X\to X$ and its conjugate $ g=h \circ f \circ h^{-1}: X' \to X'$ is related by the conjugacy $h:X \to X'$.
  If we have the iteration $f(x_i)=x_{i+1}$, under the conjugation $ h(x_i) = x_i'$, it becomes $g(x_i')=x_{i+1}'$, which is to say, one trajectory $\{x_i\}$ for the map $f(x)$ transforms into a trajectory $\{x_i'\}$ for the map $g(x')$. In particular, the cycle for the map $f(x)$ is  a cycle for the map $g(x')$. If the conjugacy $h$ is piecewise smooth, a typical trajectory has identical weight in both coordinates, which suggests that the dynamical average can be similarly computed with iterations of the map $g(x')$,
\begin{equation}
\langle a \rangle _f= \frac{1}{N} \sum_{i=i}^{N} a(x_i) = \frac{1}{N} \sum_{i=1}^N a(h^{-1}(x_i')) = \langle a \circ h^{-1} \rangle_g\,. \label{eq:cta}
\end{equation}
  Eq.~(\ref{eq:cta}) shows that  computing the average $ \langle a \rangle $ under the  map $f(x)$ is equivalent to computing the average $\langle a\circ h^{-1} \rangle$ under the map $g(x')$.

  Based on the discussion above, it is natural to  introduce a concept: the conjugate dynamical zeta function as follows.

 Suppose that the map $f(x)$ and $g(x')$ are conjugate, with $f=h^{-1}\circ g \circ h$. The observable $a$ for the map $f(x)$ has a correspondent $a\circ h^{-1}$ for the map $g(x')$. The dynamical zeta function $\frac{1}{\zeta}$ for $f(x)$ and $a$, and $\frac{1}{\zeta'} $ for $g(x')$ and $a \circ h^{-1}$ are said to be conjugate. We call $h$ the conjugacy between $ \frac{1}{\zeta}$ and $\frac{1}{\zeta'}$.

It's obvious that the dynamical averages obtained by $\frac{1}{\zeta}$ and  $\frac{1}{\zeta'} $ are the same, since $\langle a \rangle_f = \langle a \circ h^{-1} \rangle_g $. Typical forms for $\frac{1}{\zeta}$ and $\frac{1}{\zeta'}$ are
\begin{equation}
\begin{array}{lll}
\frac{1}{\zeta}& = &\prod_p (1-t_p),\,t_p=z^{n_p}\frac{e^{\beta A_p}}{|\Lambda_p|}\\
\frac{1}{\zeta'} & = & \prod_p (1-t_p'),\,t_p'=z^{n_p}\frac{e^{\beta A_p'}}{|\Lambda_p'|} \,,
\end{array}
\end{equation}
where $A_p'= \sum_{i=1}^{n_p}a(h^{-1}(x_i'))=\sum_{i=1}^{n_p}a(x_i) = A_p$. So, $A_p$ is equal to $A_p'$, no matter if $ h$ is diffeomorphic or not. If $h$ is diffeomorphic, the cycle stability eigenvalue $\Lambda_p$ is also equal to  $\Lambda_p'$. Thus, we obtain $\frac{1}{\zeta} = \frac{1}{\zeta'}$. However, when $h$ is not diffeomorphic, the cycle stability eigenvalue could  be changed and so does the dynamical zeta function, \emph{i.e.}, $\frac{1}{\zeta} \neq \frac{1}{\zeta'}$. This  happens when we use a non-diffeomorphic coordinate transformation to clear out the singularities in the natural measure.

 \subsubsection{The associated changes with the  dynamical zeta function}
 We mentioned in Section~\ref{sec:cos} that to clear out the singularities, we need a coordinate transformation $h \propto {|x-x_0|}^{1/k}$ near the singular point $x_0$. The derivative of $h$ has the form: $\frac{dh}{dx} \propto \frac{1}{{|x-x_0|}^{1-\frac{1}{k}}}$. So, the $dh/dx $ has a singularity at $x=x_0$. Thus, the conjugacy $h$ is not diffeomorphic at the singular point. If one periodic point happens to be singular, the stability eigenvalue of the periodic orbit will be changed.

 For example, for the map $f(x)=1-{|2x-1|}^k $, the fixed point $\overline{0}$ is a singular point of natural measure. So, the stability $\Lambda_0$ for $\overline{0}$ is changed to $\Lambda_0^{1/k}$, as shown below.

In the neighborhood of the fixed point $\overline{0}$, the asymptotic form for $ f $ and $ h $ is $ f \sim \Lambda_0 x,\,x > 0$ and $h \sim ax^{\frac{1}{k}}$, where $ a>0 $ is a coefficient. We have
\begin{equation}
\begin{array}{lll}
g(x')& = & h \circ f \circ h^{-1}(x') \\
 &  \sim & h \circ f(x^{k}/a^k) \sim h(\Lambda_0x^k/a^k) \sim  \Lambda_0^{1/k} x'
\end{array}
\end{equation}
 Hence stability $\Lambda_0$  of $\overline{0}$ for the conjugate map $g$ is changed to $ \Lambda_0^{1/k}$.

So, in this situation, $\frac{1}{\zeta'} \neq \frac{1}{\zeta}$. Nevertheless,  for the map $f(x)=1-{|2x-1|}^k $, if the conjugacy is diffeomorphic except at $x=0,1$, the only change in $\frac{1}{\zeta'}$ is $\Lambda_0'=\Lambda_0^{1/k}$, as compared to $\frac{1}{\zeta}$. As the conjugate dynamical zeta function $\frac{1}{\zeta'}$ is computed for map $g(x')$ whose natural measure has no singularity,  the convergence for $\frac{1}{\zeta'}$ is much improved than for $\frac{1}{\zeta}$. Note that the exact functional form of the conjugacy is not essential as long as it has the right asymptotic form near the singular point.

\subsection{Several examples}
 We  proved that  dynamical averages in a map can be computed with its conjugate map. For a map with critical points, we should find an appropriate coordinate transformation to clear out  singularities in the natural measure.  The conjugate map behaves much better in the sense that  the singularity caused by critical points is eliminated and the convergence of the conjugate dynamical zeta function is accelerated. Moreover, we do not have to know the exact functional form of the coordinate transformation. What we do is  change stability eigenvalues of specific cycles supported on the  singularities of the natural measure and hence transform the  dynamical zeta function to its conjugate.

 In the following, we apply our  method to several examples. All these maps have critical points and therefore, produce a natural measure with singularity.

\subsubsection{The logistic map}
The logistic map $f(x)=4x(1-x)\,,x\in[0,1]$  has a critical point of order two. Its natural measure is singular at two points: $x=0$ and $x=1$. Under the transformation $h(x)=\frac{2}{\pi}\arcsin{\sqrt{x}}$, the singularities in its natural measure are cleared out, as shown in FIG.~\ref{fig:10}. Also, the conjugate map $g(x')=h \circ f(x) \circ h^{-1}$ has no critical point any more, as depicted in FIG.~\ref{fig:9}.

 In the map $g(x')$, the stability eigenvalue of the fixed point $\overline{0}$ is  $\Lambda_0^{1/2}=2 $ where $\Lambda_0=4$ is the stability eigenvalue of the corresponding point in the logistic map, with stability eigenvalues of all the other orbits remaining  the same.

The logistic map has an interesting property: the eigenvalue for any prime cycle except  $\overline{0}$ has an absolute value of $ 2^{n}$, where $n$ is the length of the cycle. However, the eigenvalue of $\overline{0}$ is $\Lambda_0=4$. After the coordinate transformation, the absolute value of the eigenvalue of any prime cycle of length $n$ is $2^n$, returning to the tent map case.

Thanks to this interesting observation, the conjugate dynamical zeta function under the conjugation $h$ is just
\begin{equation}
\frac{1}{\zeta}=1-z,
\end{equation}
which is the same  dynamical zeta function of the tent map. So, the escape rate of the logistic map is exactly $0$.

\begin{figure}[htp]
\includegraphics[width=0.48\textwidth,height=0.3\textwidth]{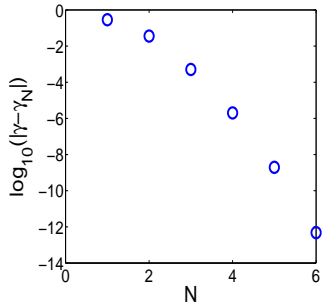}
\caption{ The error of the escape rate for the logistic map computed with the conjugate spectral determinant.}
\label{fig:11}
\end{figure}

Similar to the conjugate dynamical zeta function, we can write down
the conjugate spectral determinant, only with the change $\Lambda_0'
= \Lambda_0^{1/2} = 2$. With the conjugate spectral determinant, the
error of the escape rate decreases super-exponentially, as shown in
FIG.~\ref{fig:11}. However, if we apply the spectral determinant
directly to the logistic map, the error of the escape rate decreases only exponentially with the truncation length $N$, as shown in
FIG.~\ref{fig:3}(b). So, by an appropriate coordinate
transformation, we bring the super-exponential convergence back,
which signals that the influence of critical points wears out in the conjugate system.

Certainly, the logistic map here is very special since it is exactly
conjugate to a tent map~\cite{the95lya}. To show the general
applicability of the technique, in the following, we apply the
method to several other maps for which no smooth conjugacy to the
piecewise linear map is known.

\subsubsection{The map $f(x)=\sin(\pi x) $}
\begin{figure}[htp]
\subfigure[the map $f(x)=\sin(\pi x)$]{\includegraphics[width=0.48\textwidth,height=0.3\textwidth]{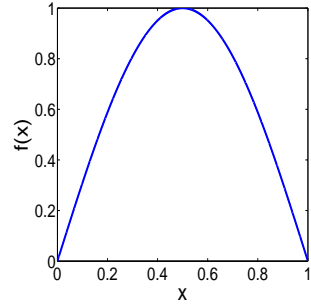}}
\subfigure[the conjugate map $g(x')$]{\includegraphics[width=0.48\textwidth,height=0.3\textwidth]{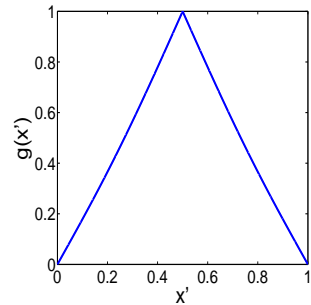}}
\caption{The graph of (a) $f(x)=\sin(\pi x)$ and (b) its conjugate map $g(x')$.}
\label{fig:12}
\end{figure}

\begin{figure}[htp]
\subfigure[the natural measure of $f(x)=\sin(\pi x)$]{\includegraphics[width=0.48\textwidth,height=0.3\textwidth]{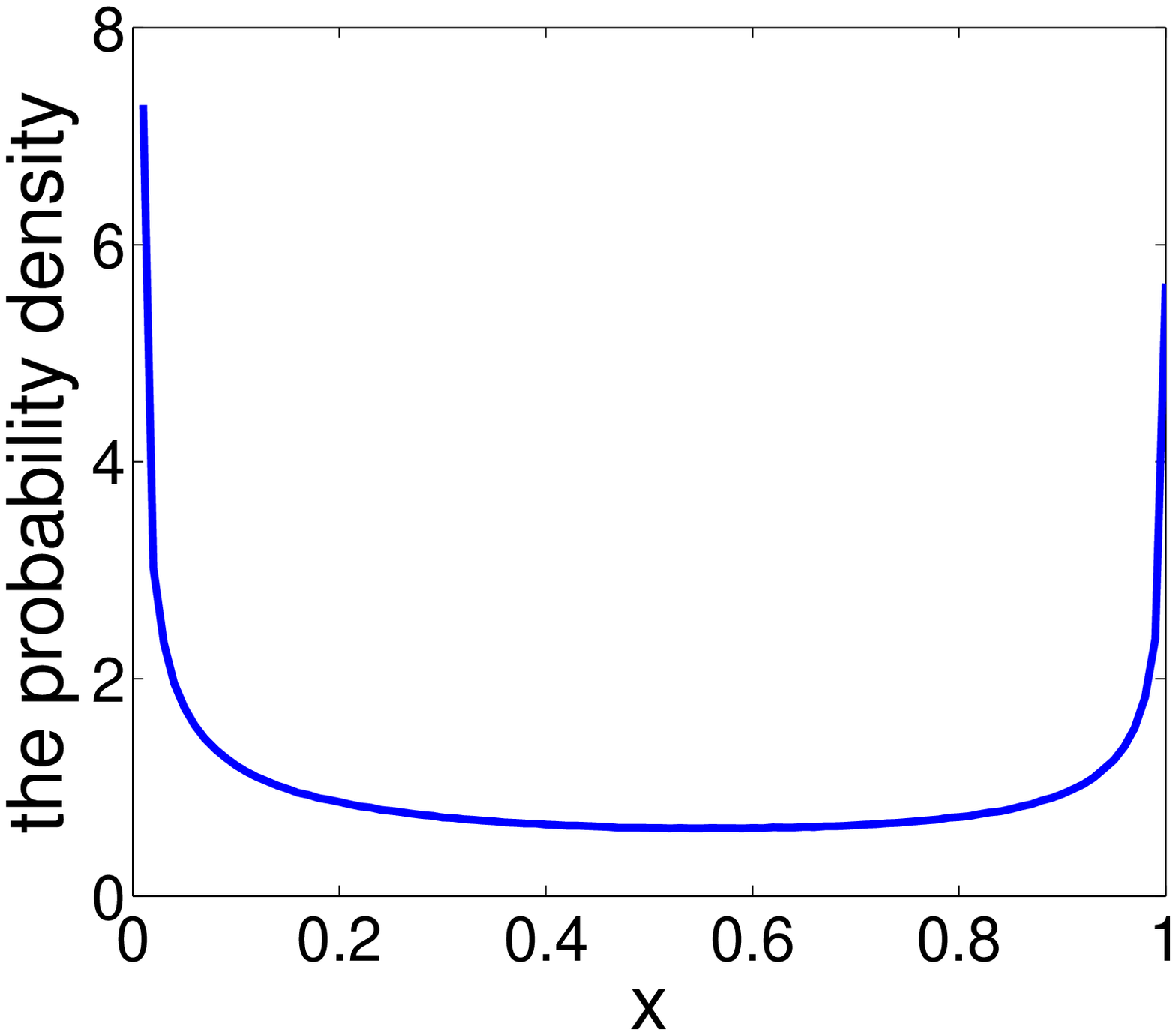}}
\subfigure[the natural measure of the conjugate map $g(x')$]{\includegraphics[width=0.48\textwidth,height=0.3\textwidth]{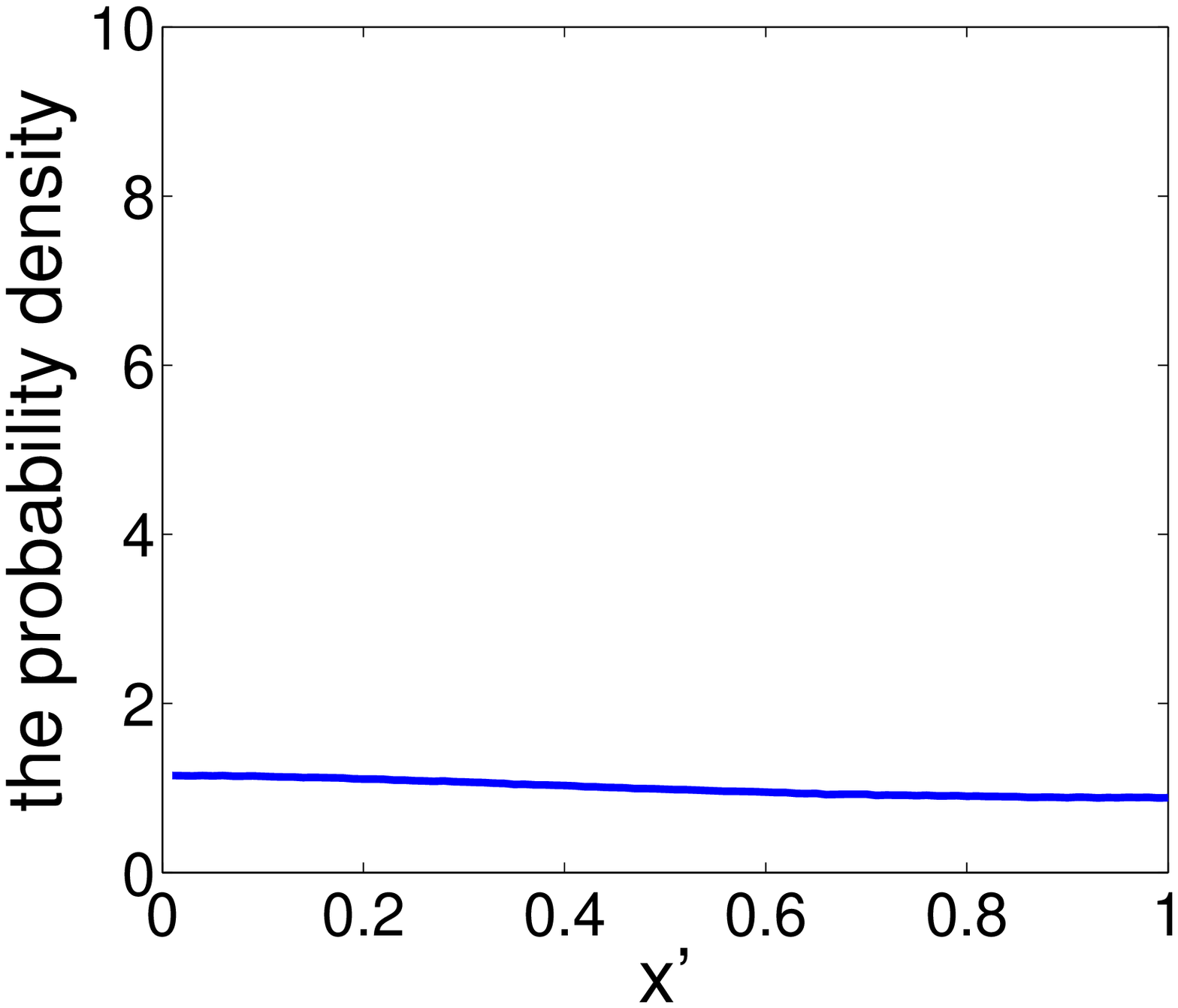}}
\caption{The natural measure of (a) $f(x)=\sin(\pi x)$ and (b) its conjugate map $g(x')$.}
\label{fig:13}
\end{figure}

The map $f(x)=\sin(\pi x)\,,x\in[0,1]$ has a critical point of order
two. Similar to the logistic map, its natural measure has two singular points: $x=0$ and $x=1$.
The asymptotic form of its natural measure near the singularity is
$\rho \sim \frac{1}{\sqrt{x}}$ near $x=0$ and $ \rho \sim
\frac{1}{\sqrt{1-x}}$ near $x=1$. Under the coordinate transformation
$h(x)=\frac{2}{\pi}\arcsin{\sqrt{x}}$, the singularities of the natural
measure are removed, and we obtain the conjugate map $g(x')=h\circ
f(x)\circ h^{-1}$. FIG.~\ref{fig:12} portraits the  map $f(x)$ and
$g(x')$ where the critical point that exists in $f(x)$ disappears in
the map $g(x')$. FIG.~\ref{fig:13} displays the natural measure
produced by the map $f(x)$ and $g(x')$ respectively where the
singularity for $f(x)$ at $x=0,1$ vanishes for $g(x')$.

For the conjugate map $g(x')$, the stability eigenvalue of the fixed point $\overline{0}$ is changed from $\Lambda_0=\pi$ to $\Lambda_0=\pi^{1/2}$, while the eigenvalue of any other orbit doesn't change. We use the dynamical zeta function and its conjugate  to calculate the escape rate, $\langle x \rangle , \,\langle x^2 \rangle, \,\langle x^3 \rangle$ of the map $f(x)=\sin(\pi x)$. The results are shown in TABLE~\ref{ta:3}, with a cutoff of cycle length $10$.  Also included are the results obtained by direct time averaging, with $10^7$  iterations. From TABLE~\ref{ta:3} we can see that the results obtained from the conjugate dynamical zeta function are by far the most accurate.
\begingroup
\squeezetable
\begin{table}[htp]
\begin{tabular}{|c|c|c|c|}
\hline
 & the dynamical  & the conjugate & time\\
 &    zeta function  & dynamical zeta function & averaging\\
\hline
escape rate & $9\times10^{-4}$ & $-3 \times 10^{-11}$ &  \\
\hline
$\langle x \rangle$ & $0.47$ & $ 0.467962949$ & $0.468$\\
\hline
 $ \langle x^2 \rangle$ & $ 0.34$ & $ 0.34397492$ & $0.344$\\
 \hline
 $\langle x^3 \rangle $ & $ 0.28$ & $ 0.28394728$ & $ 0.284$\\
 \hline
\end{tabular}
\caption{The escape rate, $\langle x \rangle\,,\langle x^2 \rangle\,,\langle x^3 \rangle$ for the map $f=\sin(\pi x) $ computed with three different methods.}
\label{ta:3}
\end{table}
\endgroup

\begin{figure}[htp]
\subfigure[ the error of the escape rate]{\includegraphics[width=0.48\textwidth,height=0.3\textwidth]{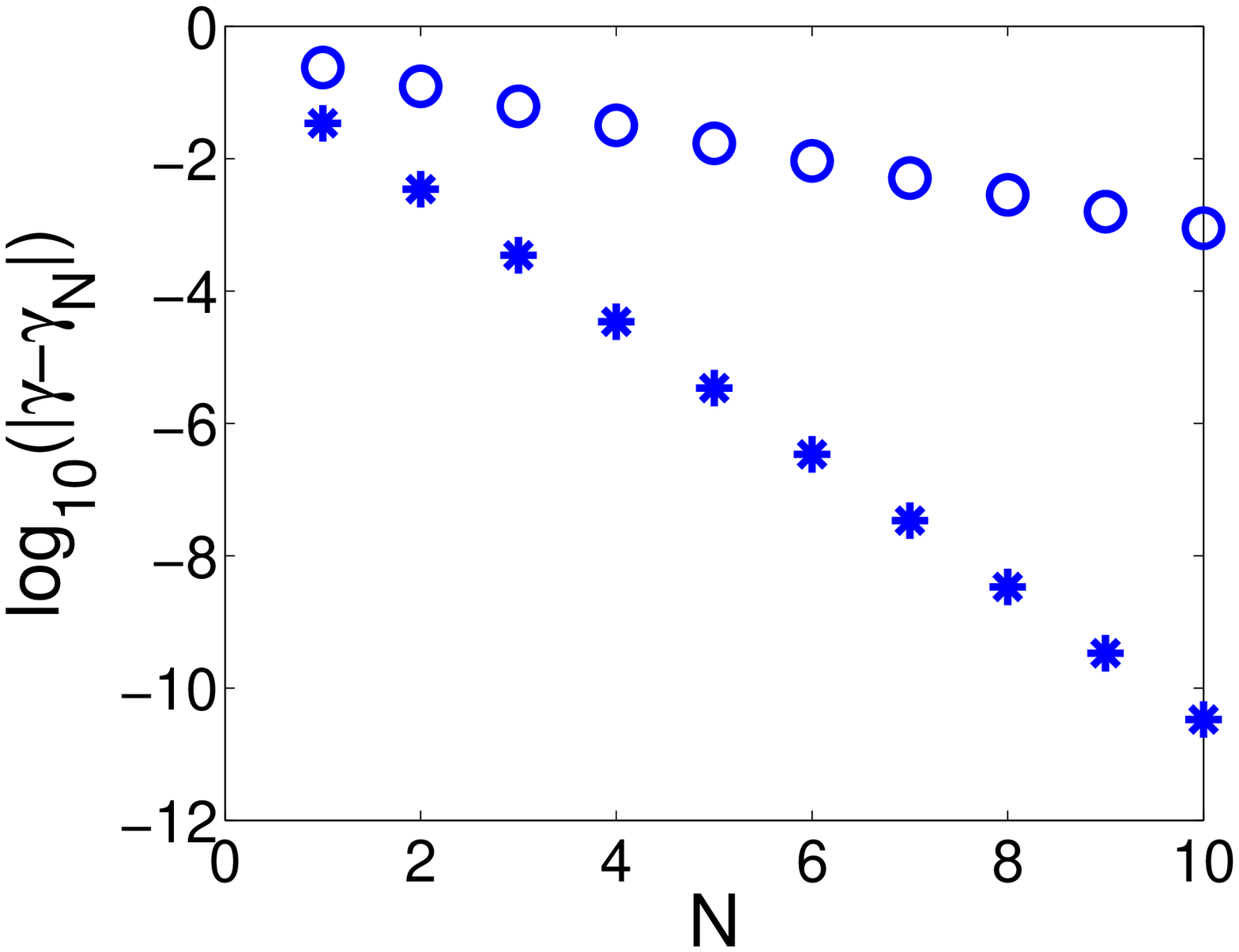}}
\subfigure[ the error of $\langle x \rangle $]{\includegraphics[width=0.48\textwidth,height=0.3\textwidth]{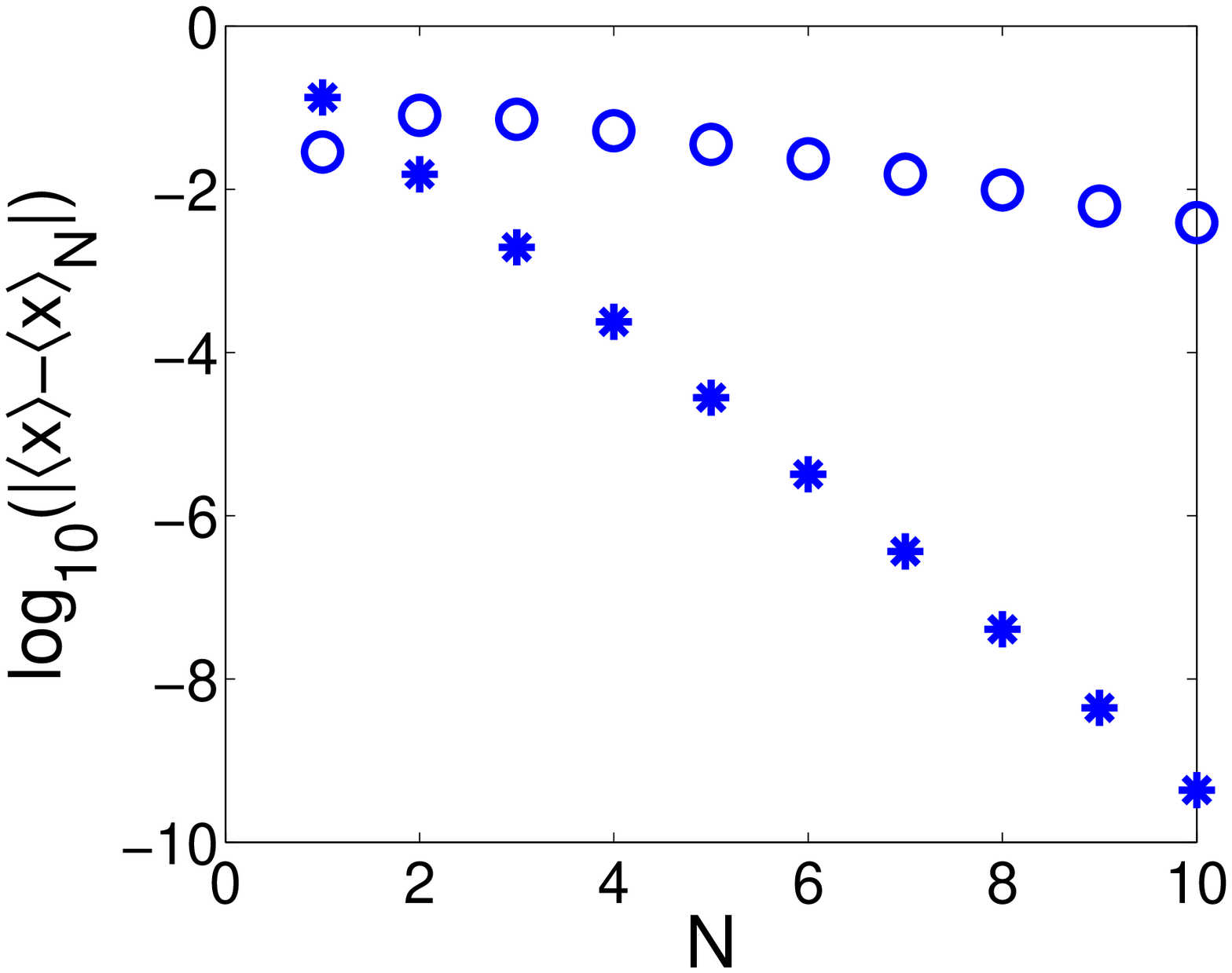}}
\subfigure[ the error of $ \langle x^2 \rangle$]{\includegraphics[width=0.48\textwidth,height=0.3\textwidth]{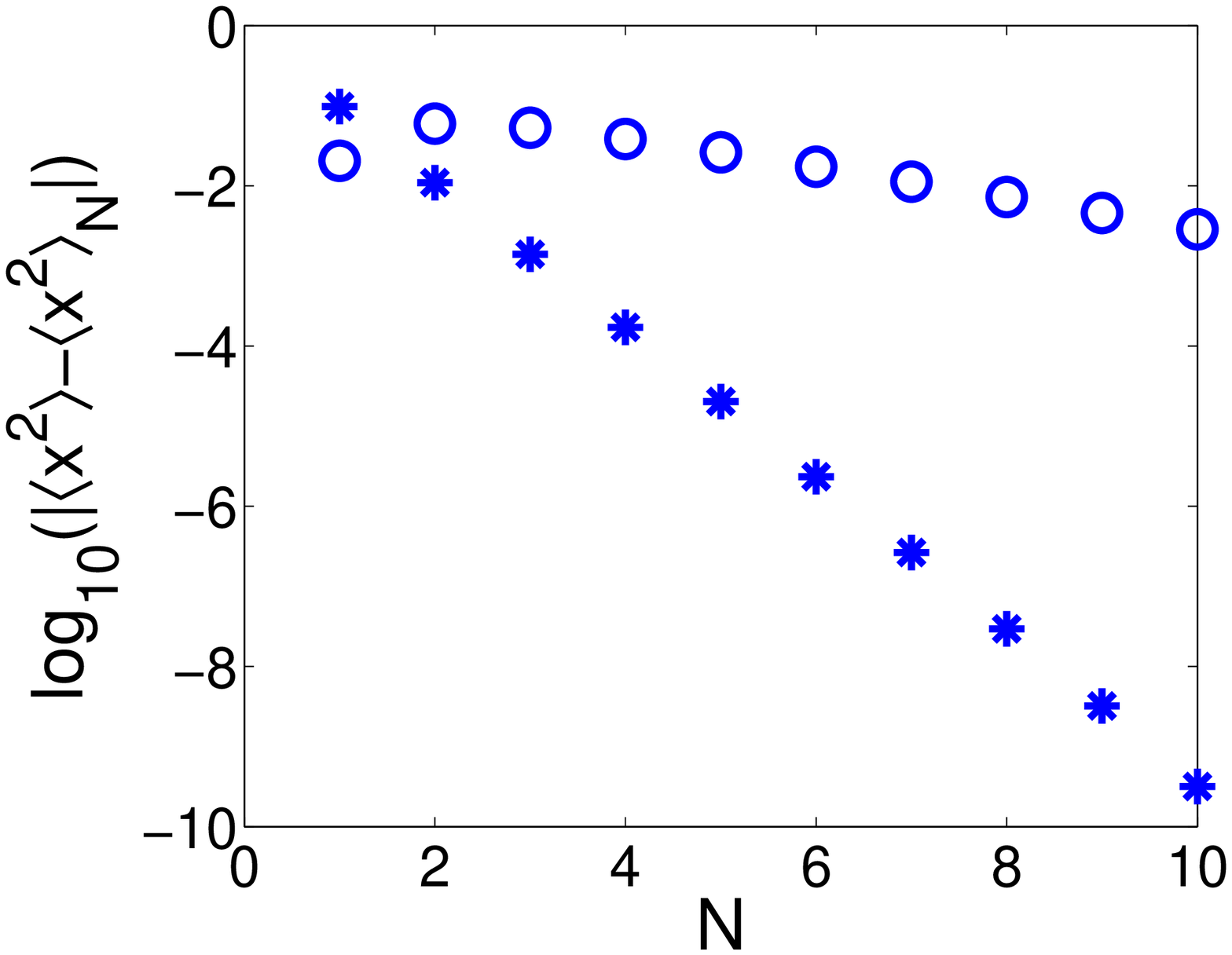}}
\subfigure[ the error of $\langle x^3 \rangle $]{\includegraphics[width=0.48\textwidth,height=0.3\textwidth]{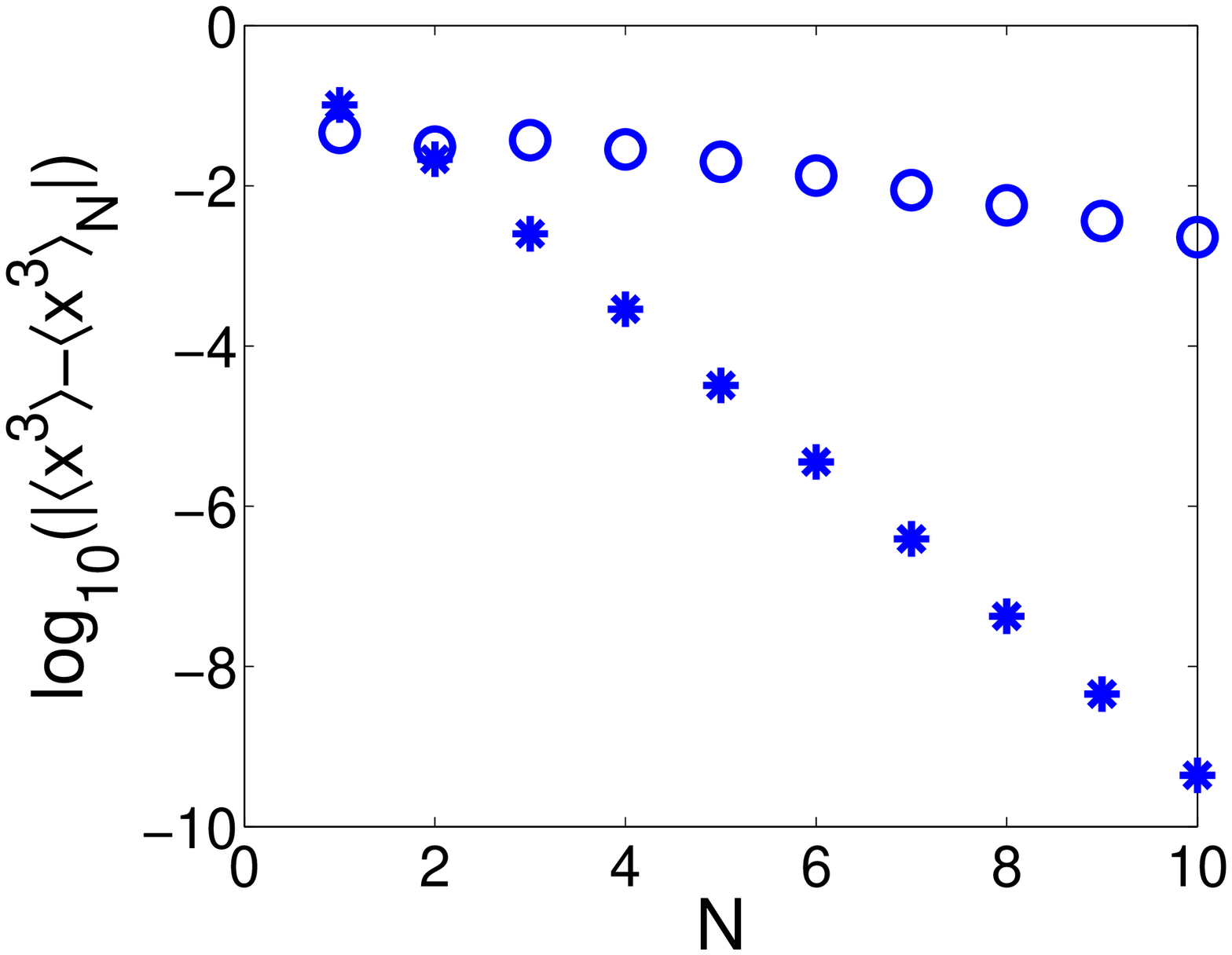}}
\caption{The error of the escape rate, $\langle x \rangle\,,\langle x^2 \rangle\,,\langle x^3 \rangle$ obtained by the dynamical zeta function(circles) for $f=\sin(\pi x)$ and its conjugate dynamical zeta function(stars).}
\label{fig:14}
\end{figure}

FIG.~\ref{fig:14} displays errors in the  averages obtained by the dynamical zeta function for $f(x)=\sin(\pi x) $ and its conjugate with different truncation length. Although all the errors seem to decrease exponentially, the results from the conjugate dynamical zeta function decay much faster, indicating a great improvement of the convergence.

\subsubsection{ The map $f(x)=1-(2x-1)^4 $}
\begin{figure}[htp]
\subfigure[the map $f(x)=1-{(2x-1)}^4 $ ]{\includegraphics[width=0.48\textwidth,height=0.3\textwidth]{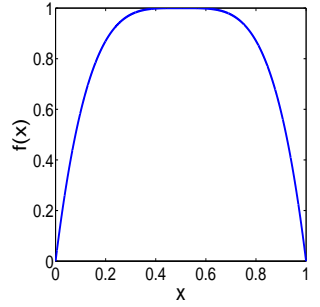}}
\subfigure[the conjugate map]{\includegraphics[width=0.48\textwidth,height=0.3\textwidth]{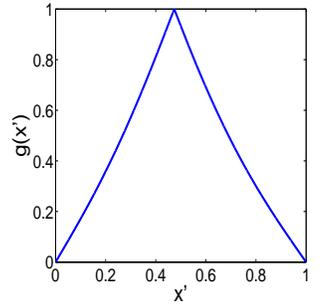}}
\caption{The graph of (a) the map $f(x)=1-{(2x-1)}^4 $ and (b) its conjugate map.}
\label{fig:15}
\end{figure}

\begin{figure}[htp]
\subfigure[the natural measure of $f(x)=1-{(2x-1)}^4 $]{\includegraphics[width=0.48\textwidth,height=0.3\textwidth]{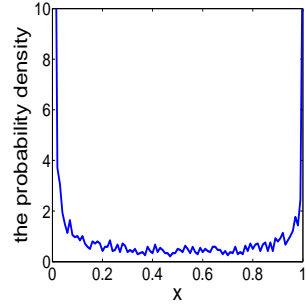}}
\subfigure[the natural measure of the conjugate map]{\includegraphics[width=0.48\textwidth,height=0.3\textwidth]{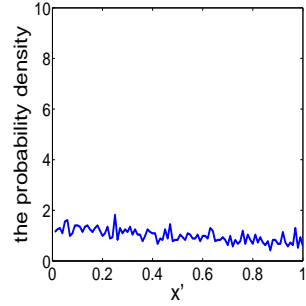}}
\caption{The natural measure of (a) $f(x)=1-{(2x-1)}^4 $ and (b) its conjugate map.}
\label{fig:16}
\end{figure}

The map $f(x)=1-(2x-1)^4$ has a critical point of order four, with measure singularities at $x=0$ and $x=1$. The asymptotic form of the natural measure near the singularity is $ \rho \sim \frac{1}{x^{\frac{3}{4}}}$ near $x=0$ and $ \rho \sim \frac{1}{(1-x)^{\frac{3}{4}}}$ near $x=1$ as shown in FIG.~\ref{fig:16}(a). To remove the singularities, an appropriate coordinate transformation is $h(x)=1-\frac{\arccos(1-2\sqrt{1-\sqrt{x}})}{\pi}$, which has an asymptotic form $ h \propto x^{1/4}$ near $x=0$ and $ h \propto (1-x)^{1/4}$ near $ x=1$. Thus, we obtain the conjugate map $g(x')=h \circ f(x) \circ h^{-1}$. The map $f(x)$ and its conjugate $g(x')$ are depicted in FIG.~\ref{fig:15}. Although $f(x)$ has a very flat top, the peak of $g(x')$ is acute. As a result, the natural measure of the map $g(x')$ has no singularity, as exhibited in FIG.~\ref{fig:16}(b).

\begin{figure}[htp]
 \subfigure[ the error of the escape rate]{\includegraphics[width=0.48\textwidth,height=0.3\textwidth]{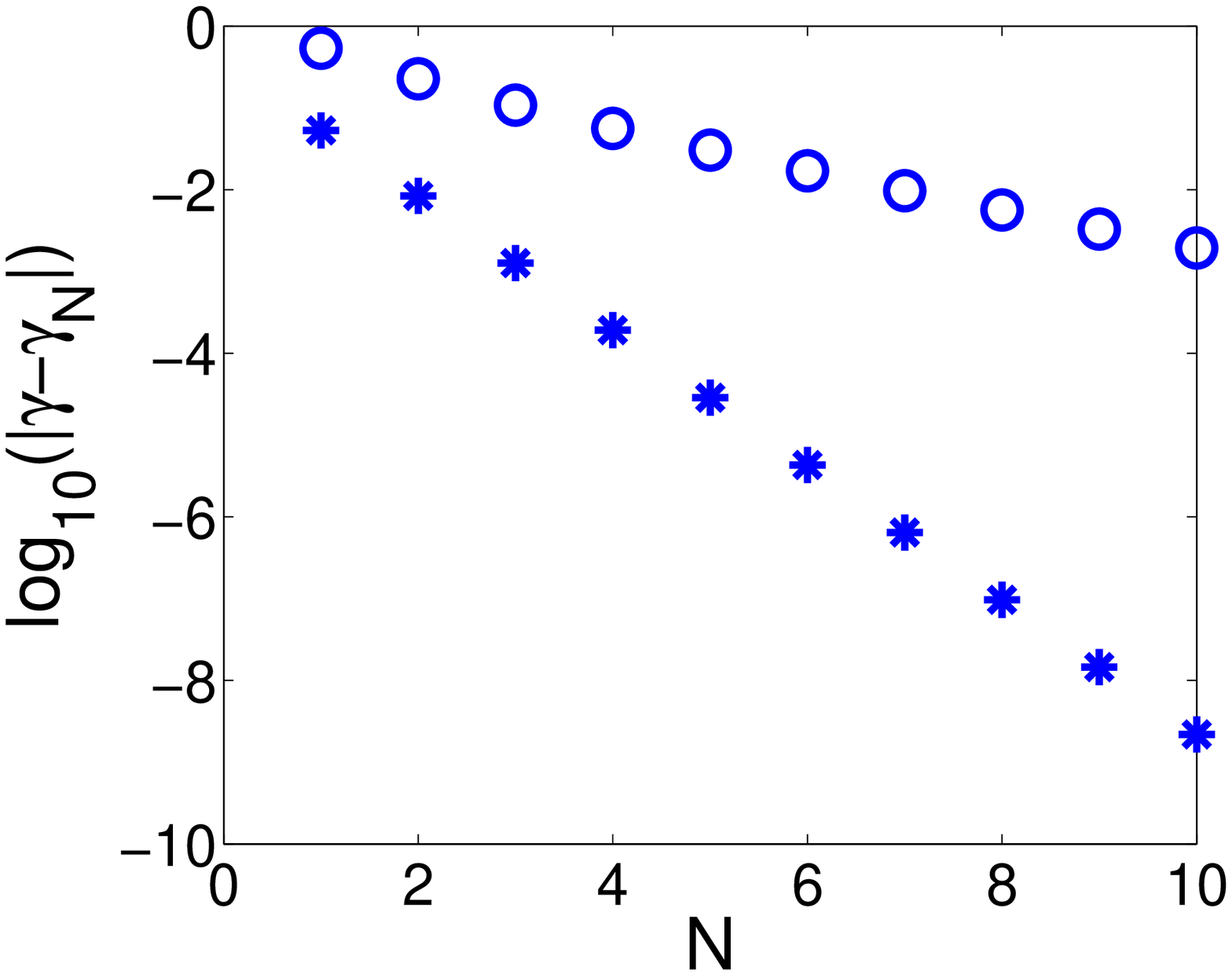}}
 \subfigure[ the error of $\langle x \rangle$]{\includegraphics[width=0.48\textwidth,height=0.3\textwidth]{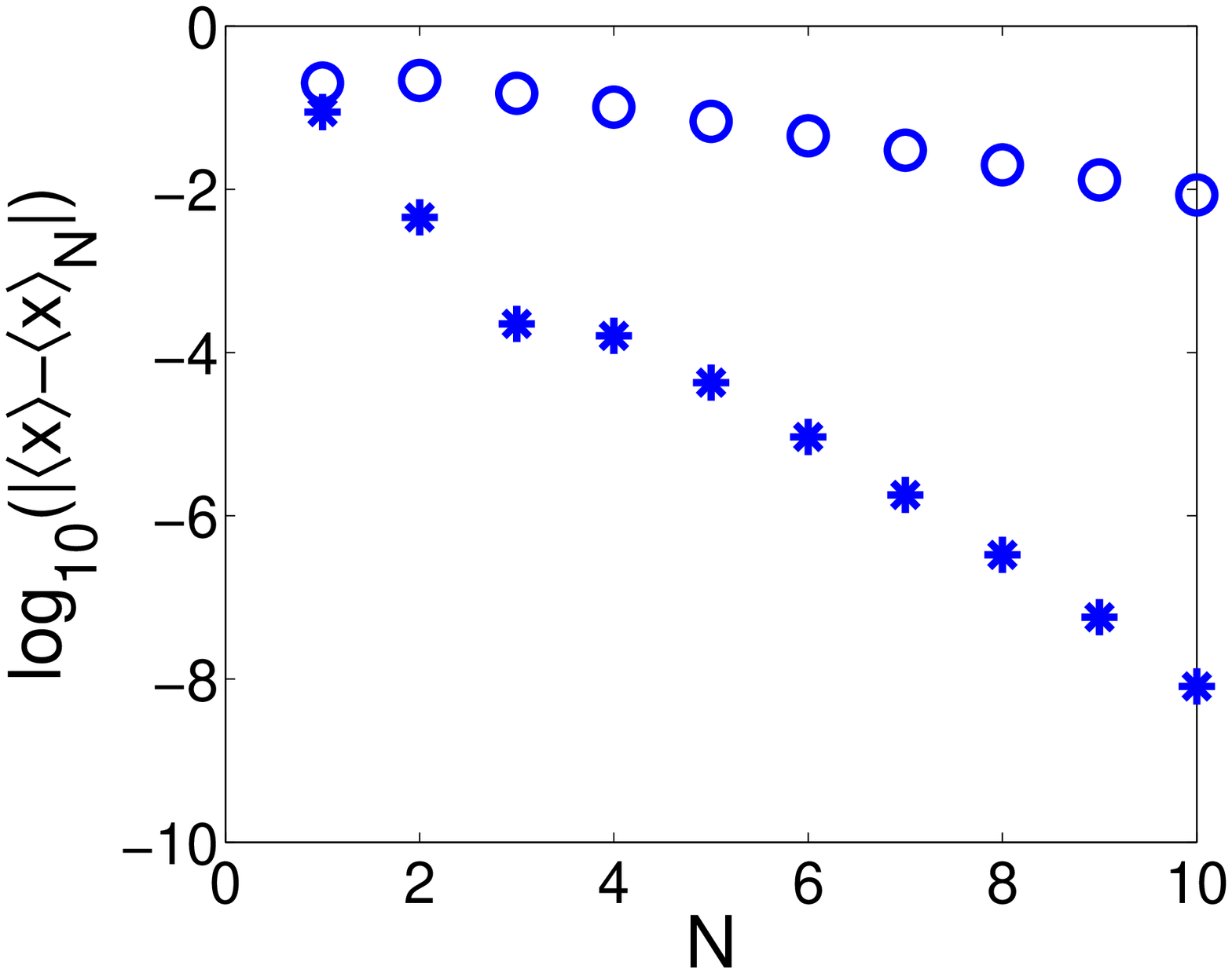}}
 \subfigure[ the error of $\langle x^2 \rangle $]{\includegraphics[width=0.48\textwidth,height=0.3\textwidth]{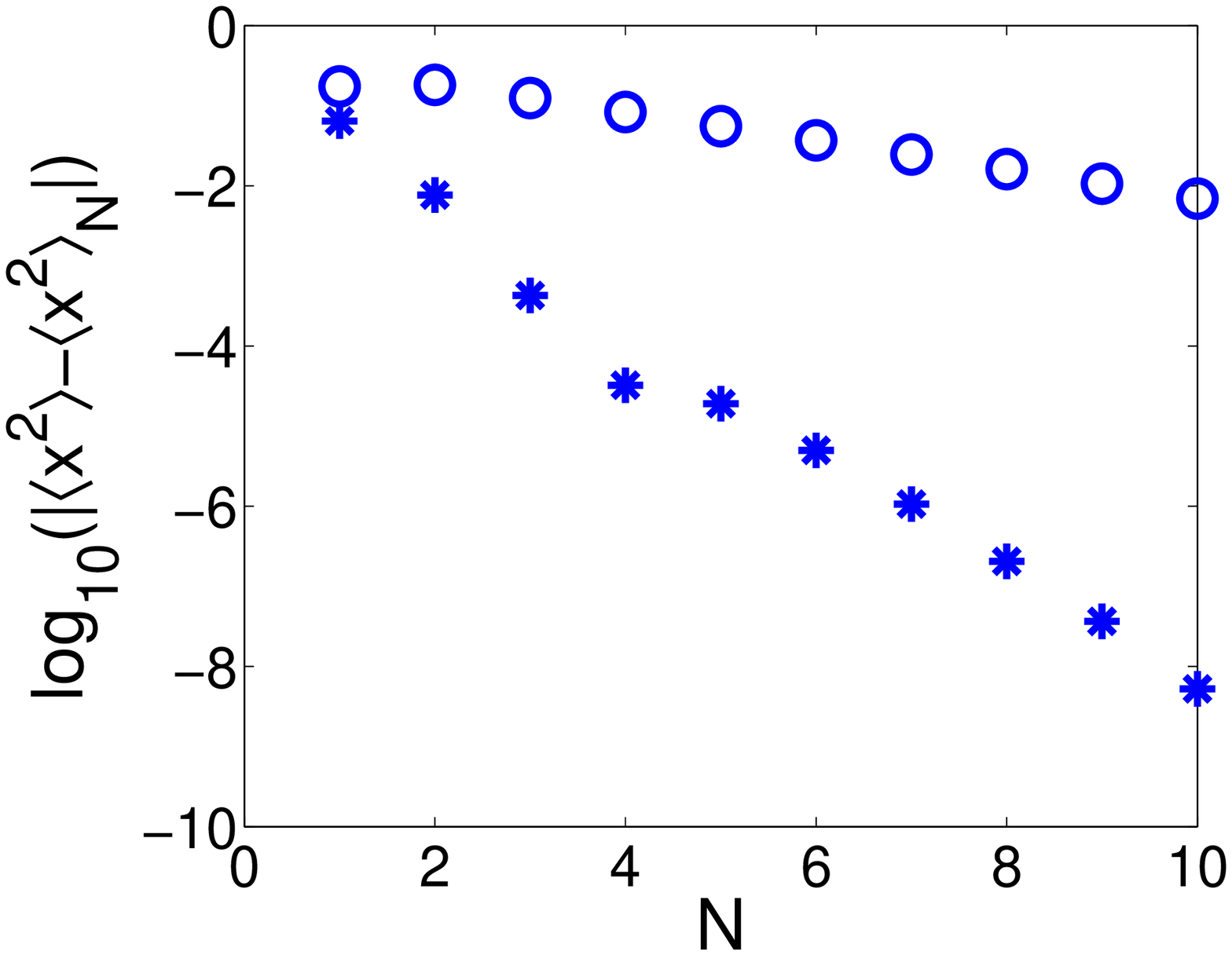}}
 \subfigure[ the error of $\langle x^3 \rangle$]{\includegraphics[width=0.48\textwidth,height=0.3\textwidth]{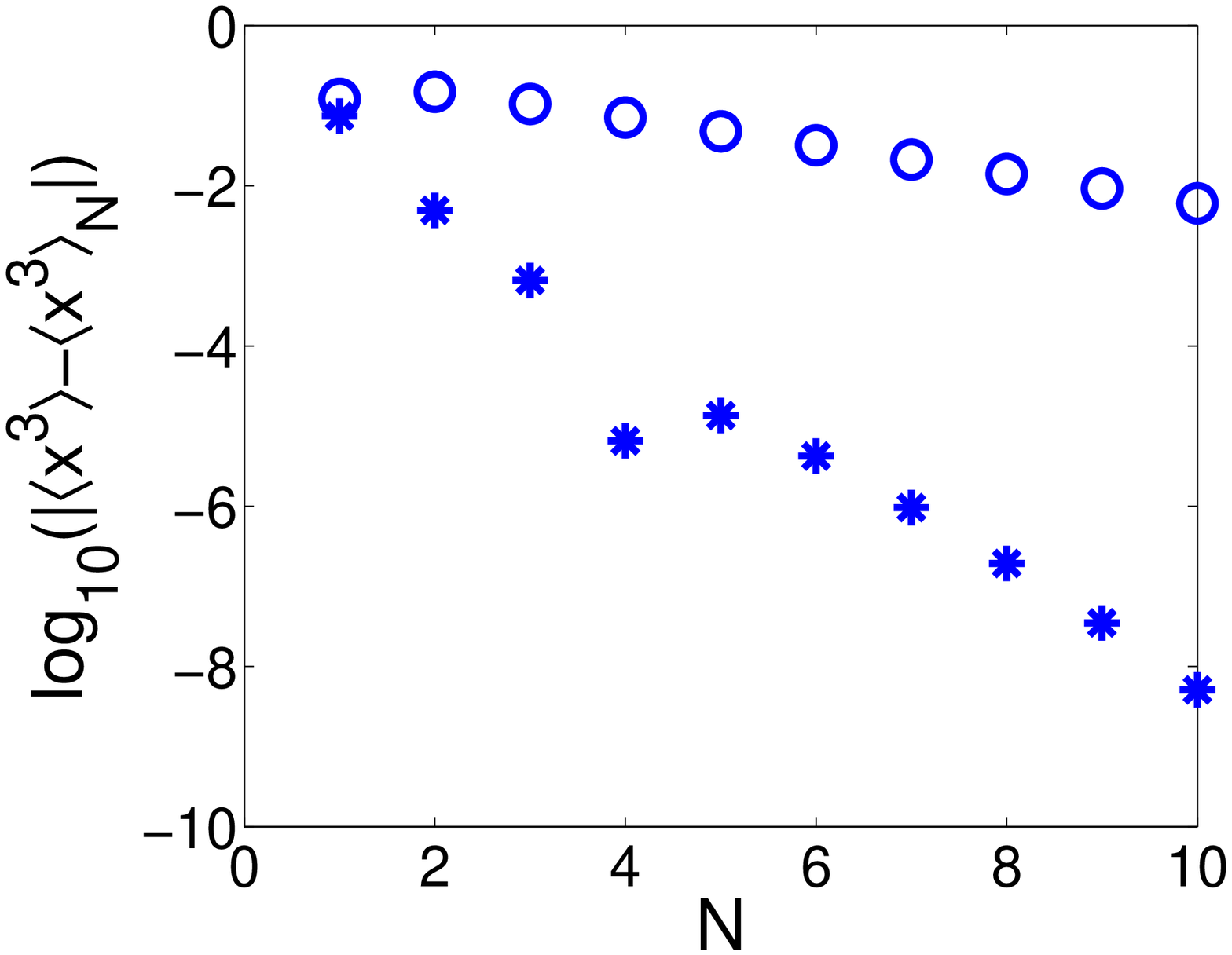}}
 \caption{The error of the escape rate, $\langle x \rangle\,,\langle x^2 \rangle\,,\langle x^3 \rangle$ obtained by the dynamical zeta function(circles) for $f(x)=1-{(2x-1)}^4$ and its conjugate dynamical zeta function(stars).}
 \label{fig:17}
 \end{figure}

The stability eigenvalue of $\overline{0}$ of the conjugate map is $\Lambda_0'=\Lambda_0^{1/4}= 8^{1/4}$. The convergence of the dynamical zeta function for $f(x)=1-{(2x-1)}^4$ is even poorer than the logistic map. However, the conjugate dynamical zeta function continues to give a  much accelerated convergence. The  escape rate, $\langle x \rangle\,,\langle x^2 \rangle\,,\langle x^3 \rangle$ obtained by the two different ways are shown in TABLE~\ref{ta:4}, with a truncation length $10$. It is worth mentioning that the direct time averaging becomes very unreliable in the current case. FIG.~\ref{fig:17} plots the errors in these averages with different truncation length.  We can see that the conjugate dynamical zeta function converges much faster.

\begin{table}[htp]
\begin{tabular}{|c|c|c|}
\hline
 & the dynamical & the conjugate \\
 & zeta function & dynamical zeta function\\
 \hline
 escape rate & $2\times10^{-3}$ & $-2\times10^{-9}$   \\
 \hline
 $\langle x \rangle $ & $0.45$ & $0.4475860$ \\
 \hline
 $\langle x^2 \rangle $ & $0.36$ & $0.3601271$\\
 \hline
 $\langle x^3 \rangle $ & $0.32 $ & $0.31801265$\\
 \hline
\end{tabular}
\caption{The escape rate, $\langle x \rangle\,,\langle x^2 \rangle\,,\langle x^3 \rangle$ for the map $f(x)=1-{(2x-1)}^4$ computed with two different methods.}
\label{ta:4}
\end{table}

\subsubsection{The map $f(x)=1-{(2x-1)}^6$}
\begin{figure}[htp]
\subfigure[the map $f(x)=1-{(2x-1)}^6 $]{\includegraphics[width=0.48\textwidth,height=0.3\textwidth]{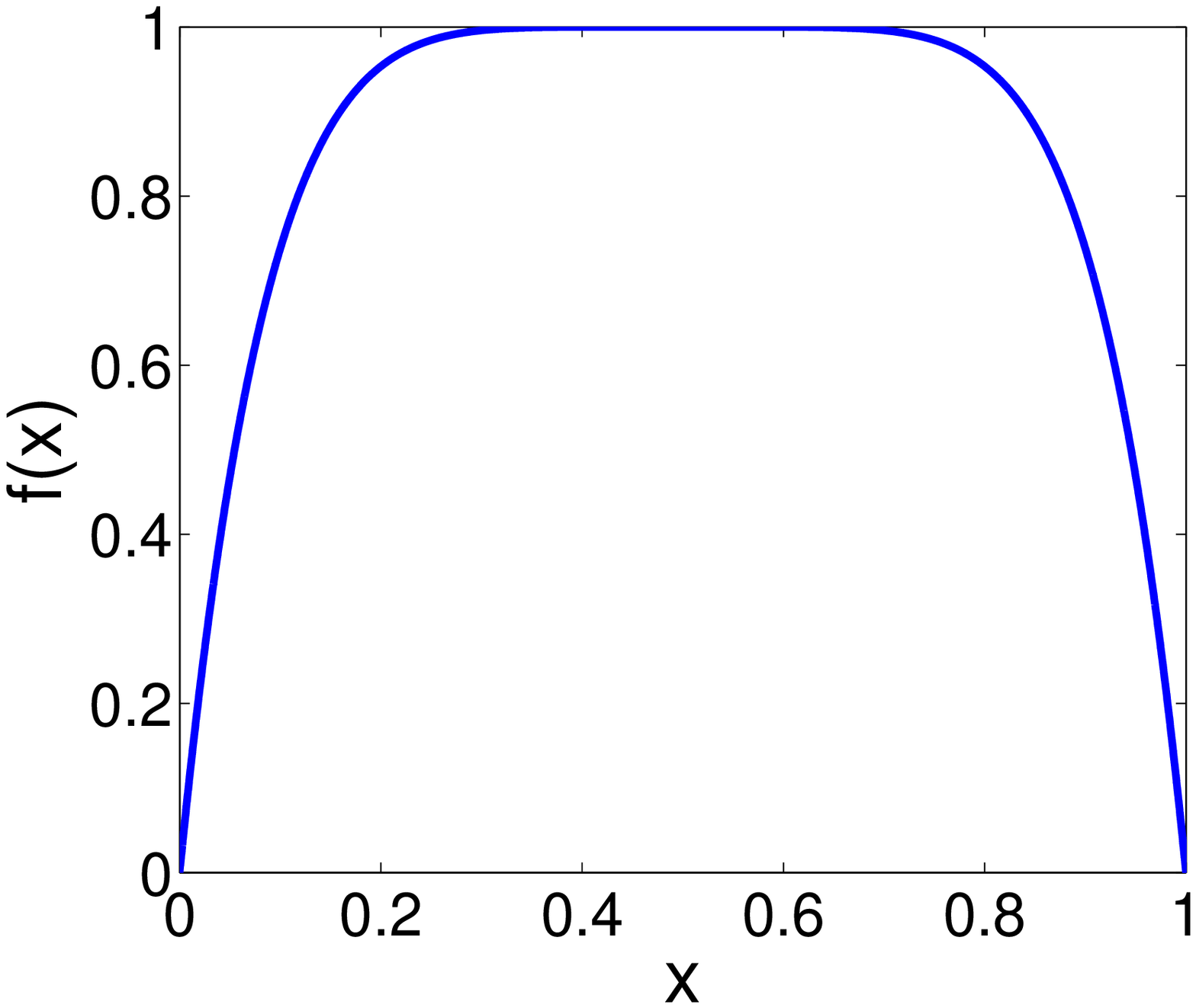}}
\subfigure[the conjugate map]{\includegraphics[width=0.48\textwidth,height=0.3\textwidth]{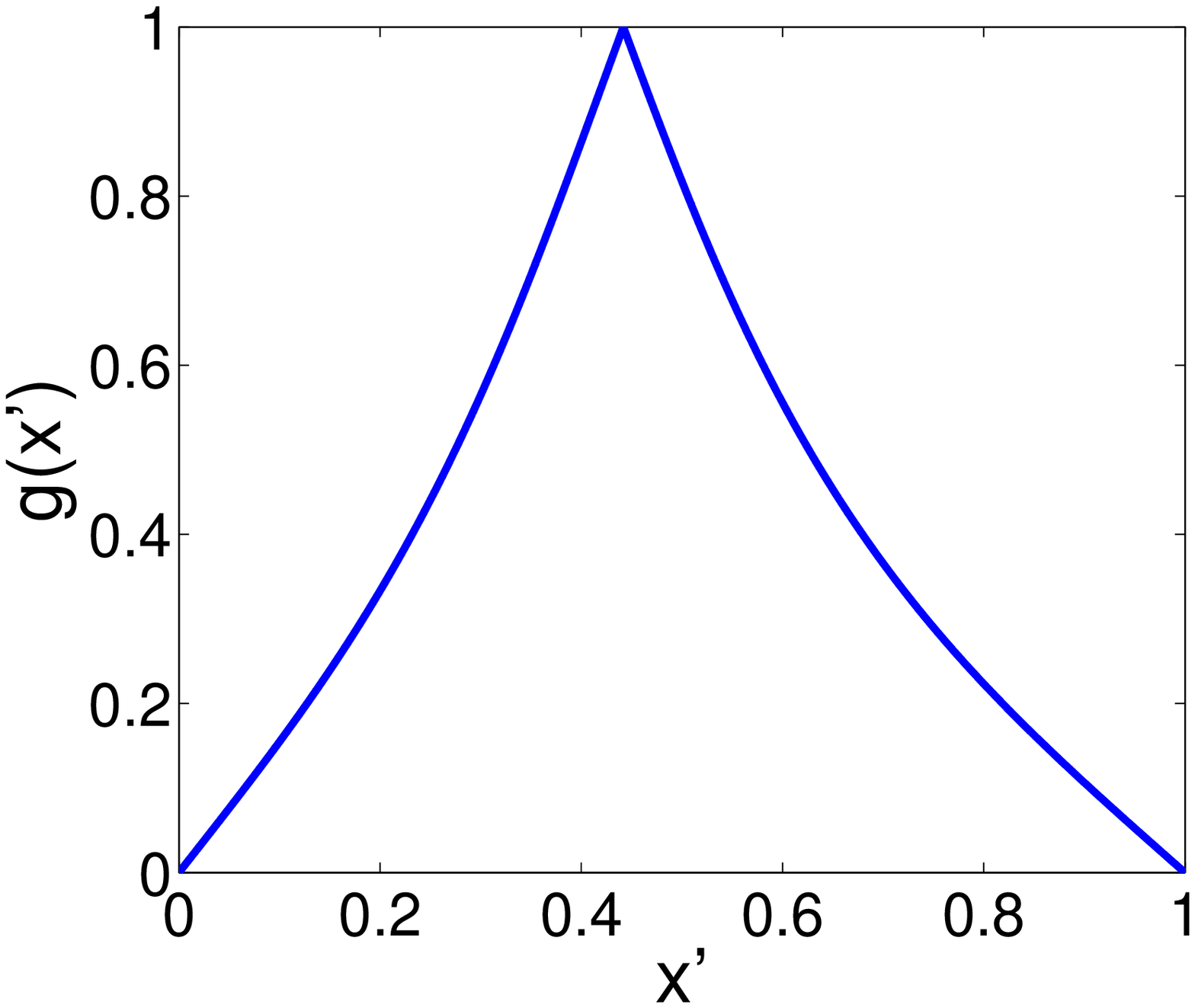}}
\caption{The graph of (a) the map $f(x)=1-{(2x-1)}^6$ and (b) its conjugate map $g(x')$.}
\label{fig:18}
\end{figure}

\begin{figure}[htp]
\subfigure[the natural measure of the map $f(x)=1-{(2x-1)}^6$]{\includegraphics[width=0.48\textwidth,height=0.3\textwidth]{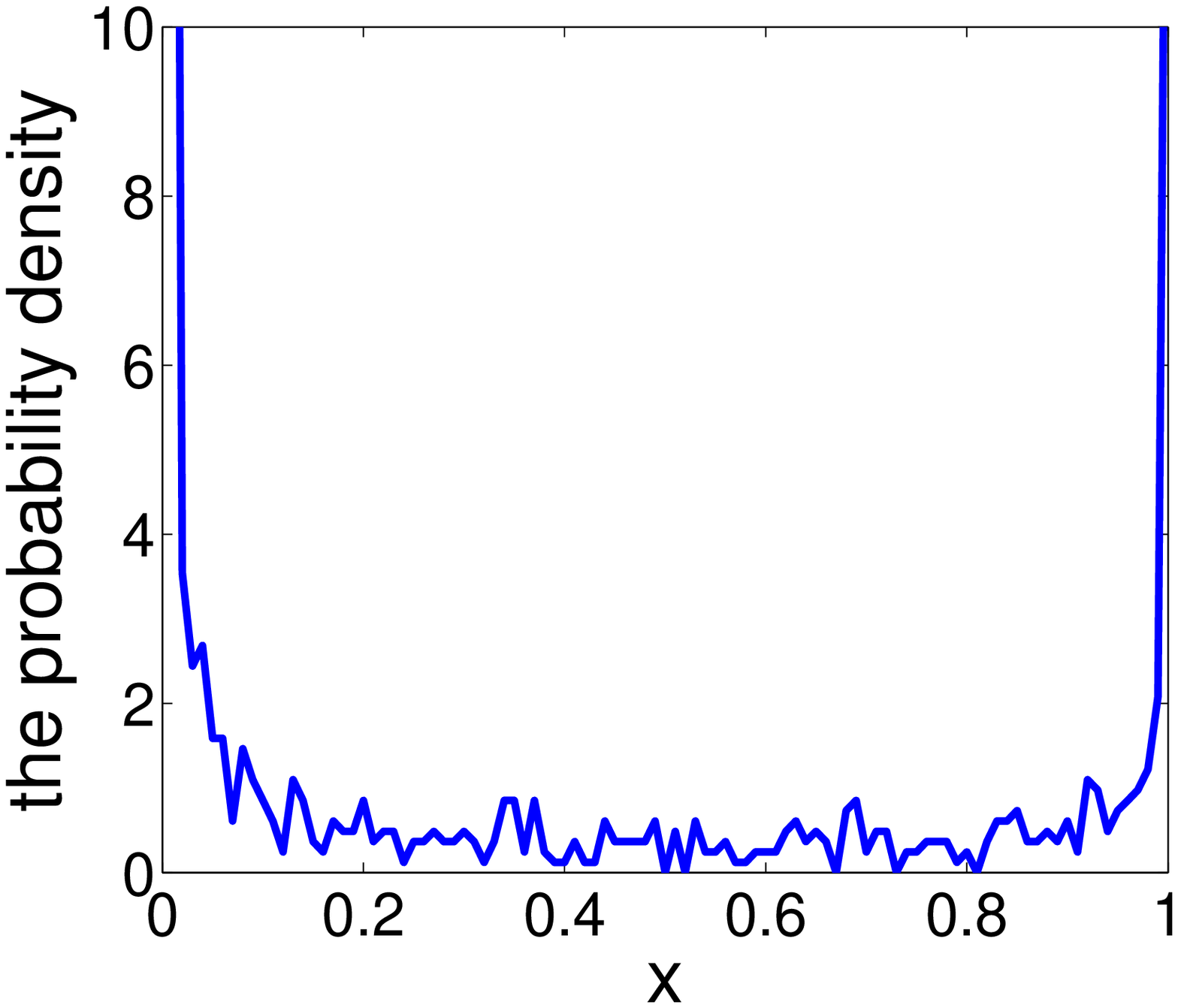}}
\subfigure[the natural measure of the conjugate map $g(x')$]{\includegraphics[width=0.48\textwidth,height=0.3\textwidth]{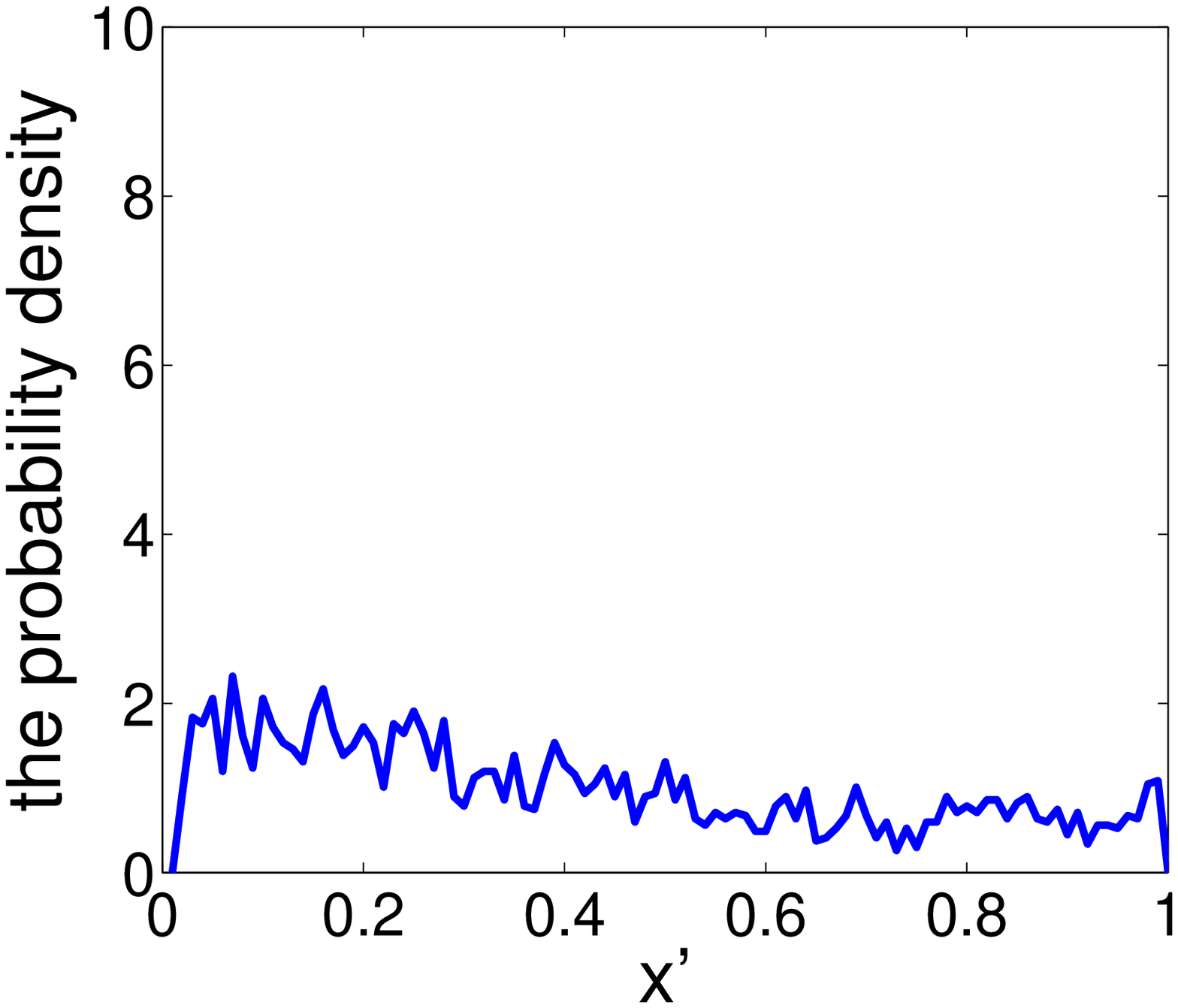}}
\caption{The natural measure of the map $f(x)=1-{(2x-1)}^6$ and its conjugate map $g(x')$.}
\label{fig:19}
\end{figure}

The map $f(x)=1-{(2x-1)}^6$ has a critical point of order six, and therefore causes an even worse convergence for the dynamical zeta function. The coordinate transformation we use is $h(x)=1-\frac{\arccos(1-2{(1-x^{\frac{1}{3}})}^{\frac{1}{3}})}{\pi} $, and the conjugate map $g(x')=h \circ f(x) \circ h^{-1} $ has no critical point any more. Thus, the singularities of natural measure are removed. FIG.~\ref{fig:18}  shows the graph of the map $f(x)$ and $g(x')$. The natural measure of the map $f(x)$ and $g(x')$ is depicted in FIG.~\ref{fig:19}, obtained with $10^8$ iterations in this case. We see that both numerical measures fluctuate, which implies that time averages based on iterations can't reach a high accuracy for this map.

\begin{figure}[htp]
\subfigure[the  error of the escape rate]{\includegraphics[width=0.48\textwidth,height=0.3\textwidth]{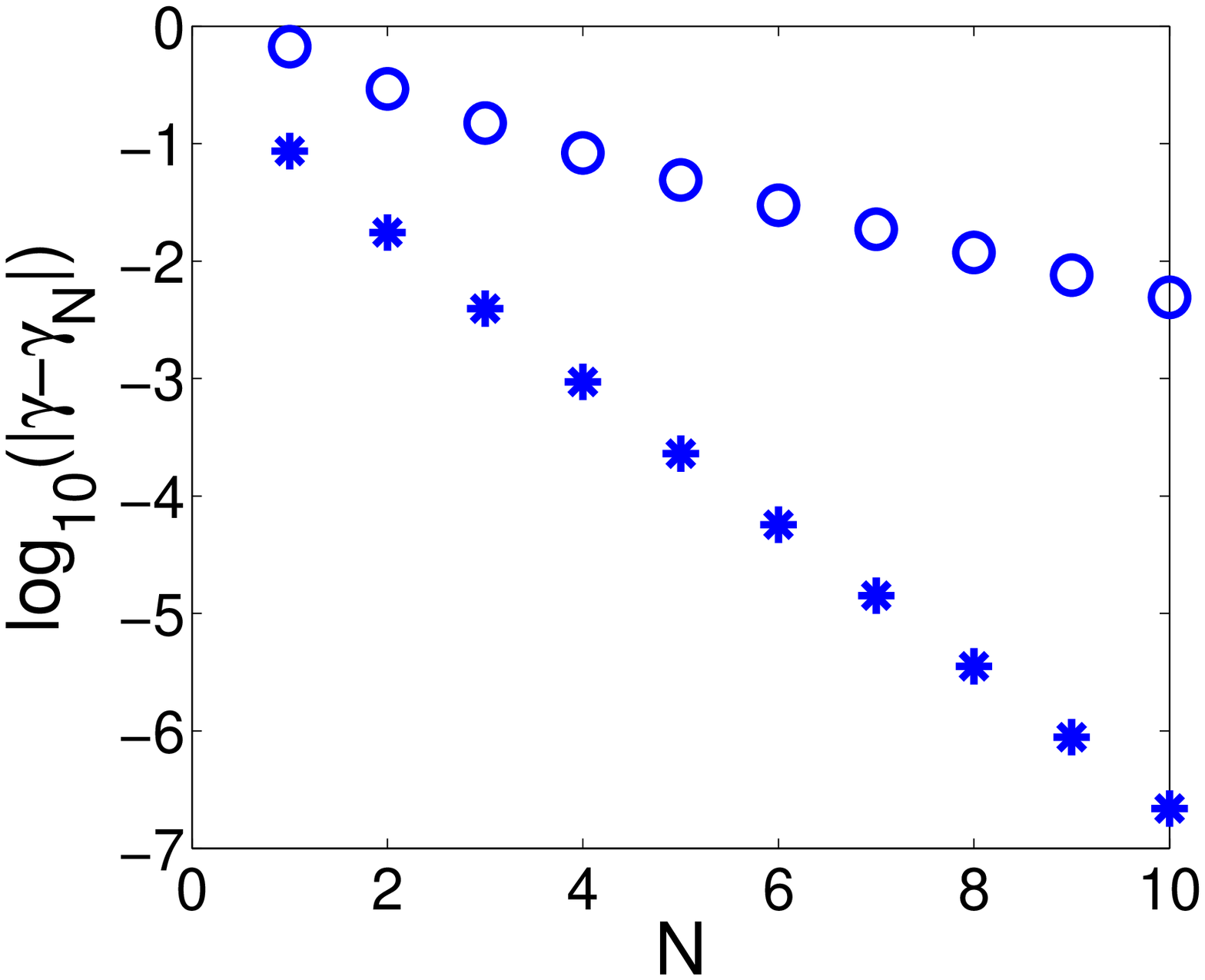}}
\subfigure[the  error of $\langle x \rangle$]{\includegraphics[width=0.48\textwidth,height=0.3\textwidth]{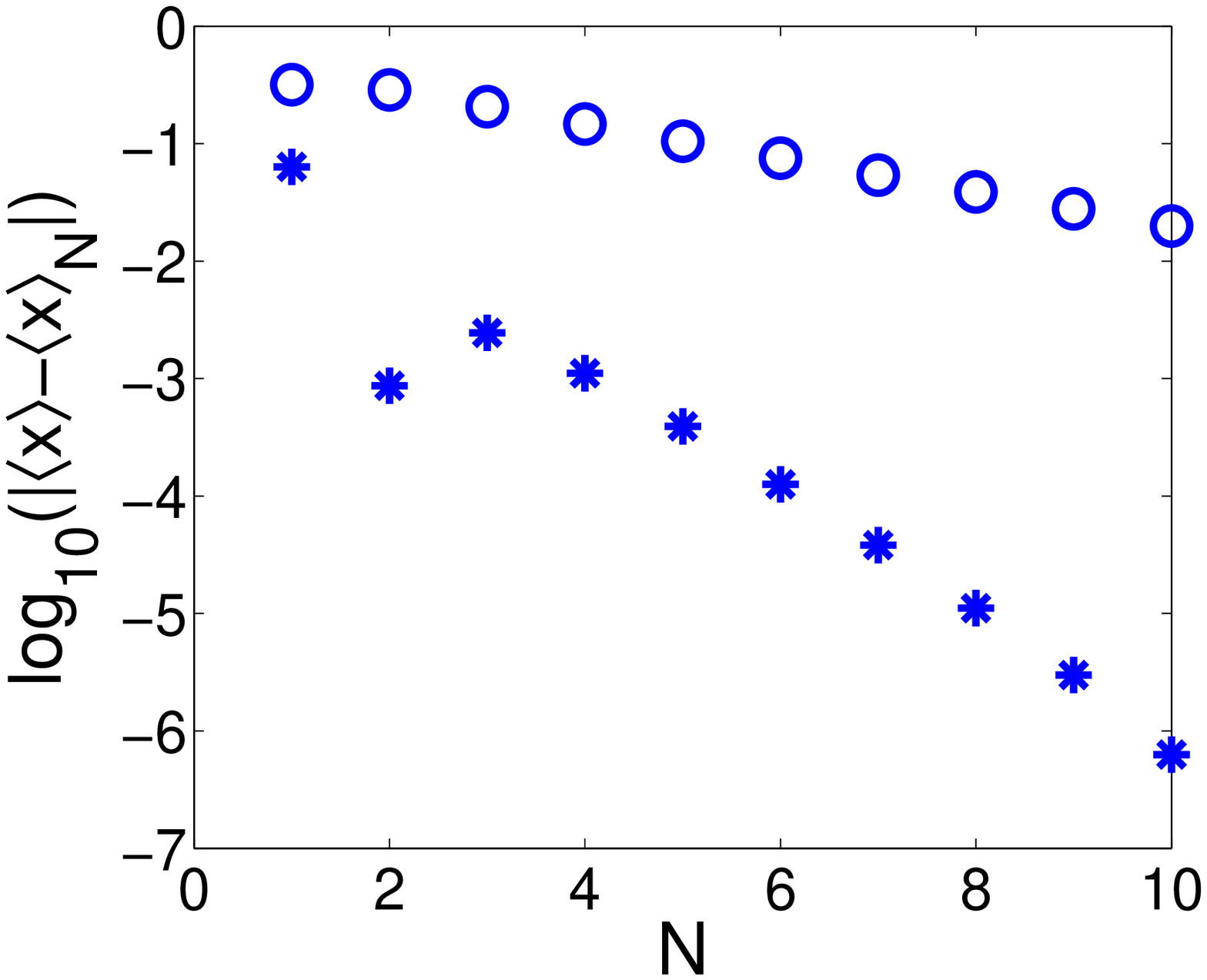}}
\subfigure[the  error of $\langle x^2 \rangle$]{\includegraphics[width=0.48\textwidth,height=0.3\textwidth]{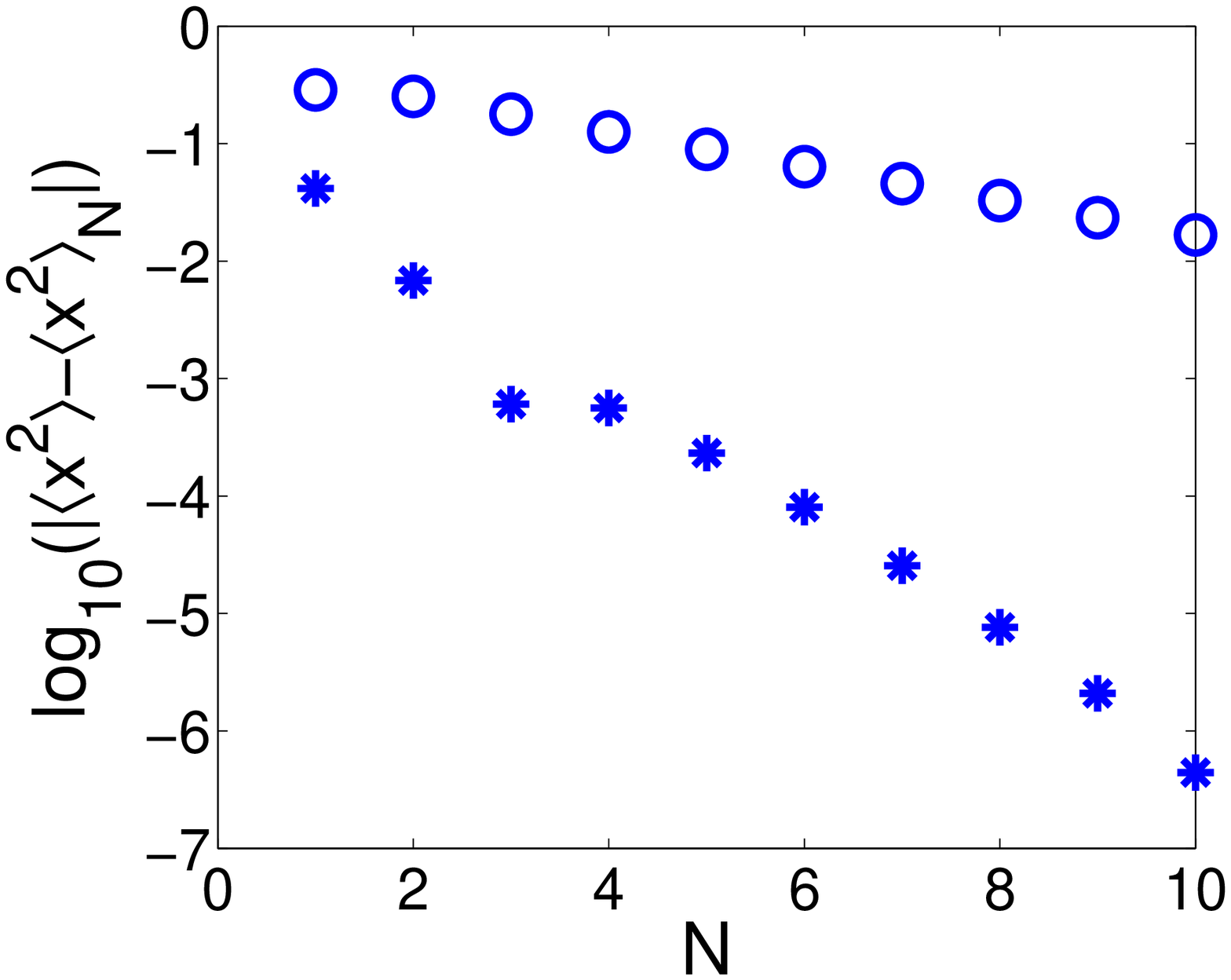}}
\subfigure[the error of $\langle x^3 \rangle$]{\includegraphics[width=0.48\textwidth,height=0.3\textwidth]{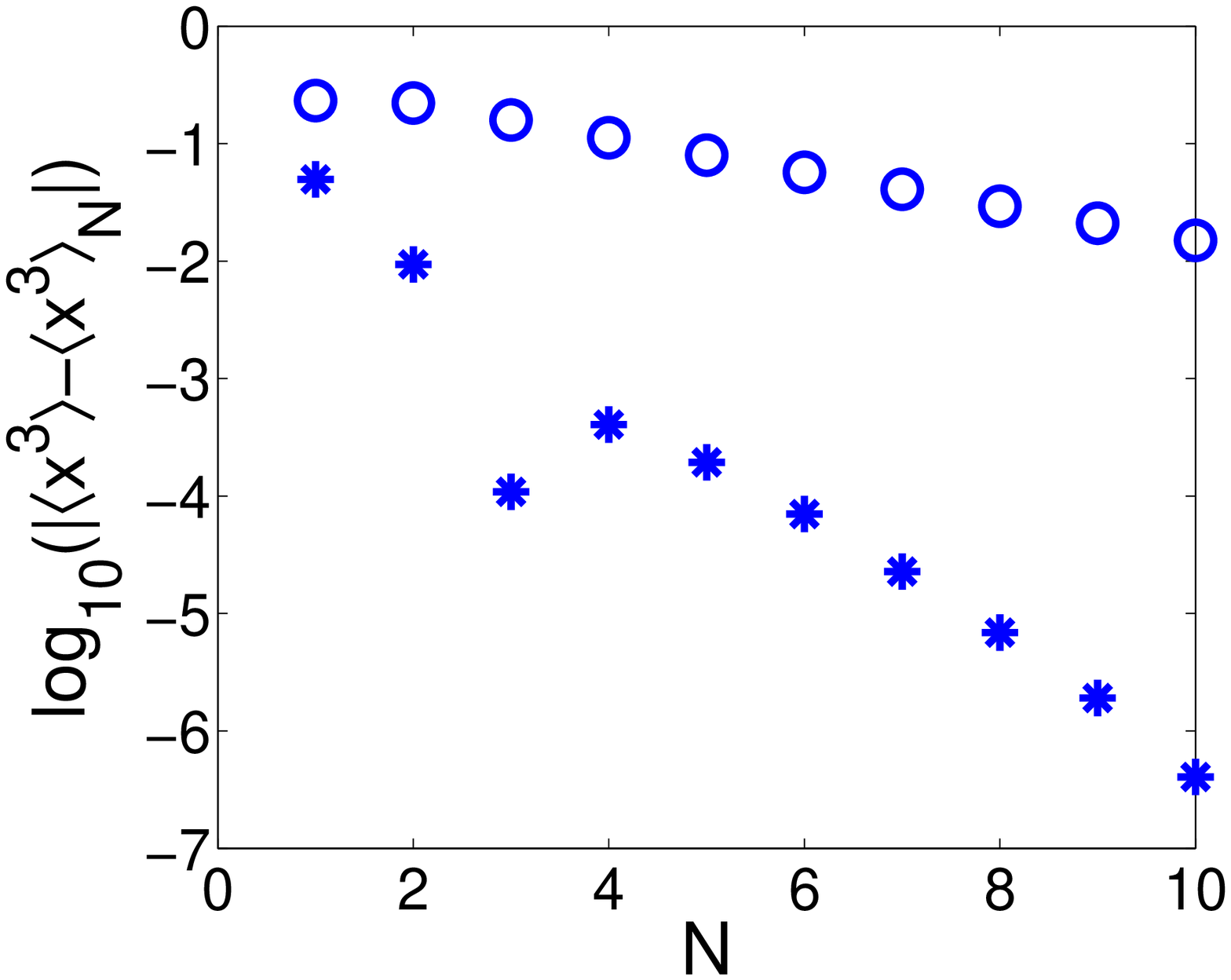}}
\caption{The  error of the escape rate, $\langle x \rangle\,,\langle x^2 \rangle\,,\langle x^3 \rangle$ obtained by the dynamical zeta function(circles) for $f(x)=1-{(2x-1)}^6$ and its conjugate dynamical zeta function(stars).}
\label{fig:20}
\end{figure}

For the conjugate dynamical zeta function, the only difference from the original one is that, the stability eigenvalue of $\overline{0}$ is changed to $\Lambda_0'=\Lambda_0^{\frac{1}{6}}=12^{\frac{1}{6}}$. The values of the averages obtained by the two dynamical zeta functions, with a truncation length $10$, are listed in TABLE~\ref{ta:5}, while the dependence of  the computational errors on the truncation length is portrayed in FIG.~\ref{fig:20}. We can see that the conjugate dynamical zeta function converges much faster than the original zeta function, which provides evidence that clearing out the singularity in natural measure helps accelerate the convergence.

\begin{table}[htp]
\begin{tabular}{|c|c|c|}
\hline
 & the dynamical  & the conjugate \\
 &  zeta function & dynamical zeta function\\
\hline
escape rate & $5 \times 10^{-3}$ & $-2 \times 10^{-7}$ \\
\hline
$\langle x  \rangle$ & $ 0.4$ & $0.40232$ \\
\hline
$\langle x^2 \rangle$ & $0.34$ & $0.332921$ \\
\hline
$\langle x^3 \rangle$ & $0.31$ & $0.30027$\\
\hline
\end{tabular}
\caption{The escape rate, $\langle x \rangle\,,\langle x^2 \rangle\,,\langle x^3 \rangle$ for the map $f(x)=1-{(2x-1)}^6$ computed with two different methods.}
\label{ta:5}
\end{table}

\subsubsection{A map with three measure singularities}
\begin{figure}[htp]
\subfigure[the map with three measure singularities]{\includegraphics[width=0.48\textwidth,height=0.3\textwidth]{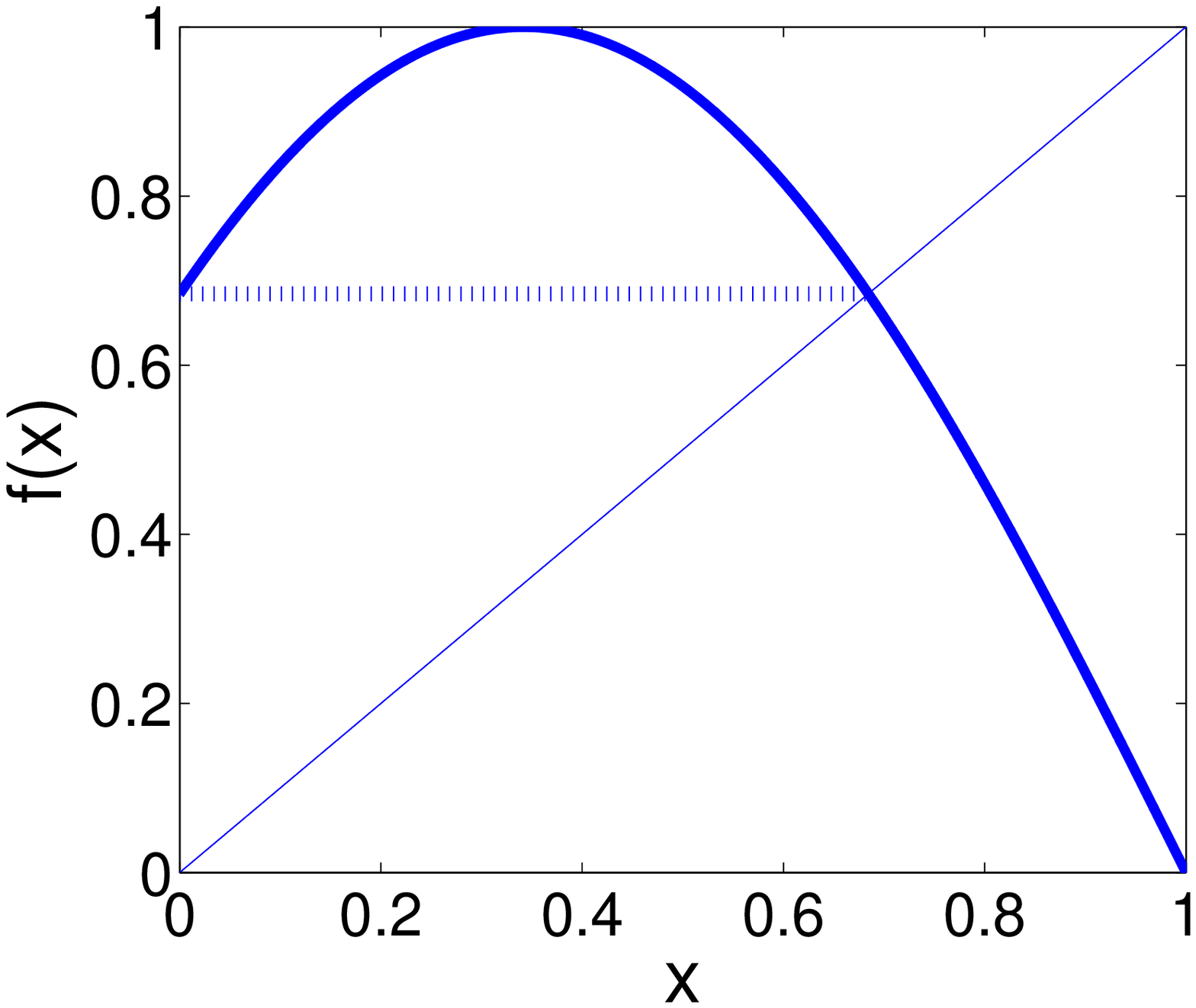}}
\subfigure[the conjugate map]{\includegraphics[width=0.48\textwidth,height=0.3\textwidth]{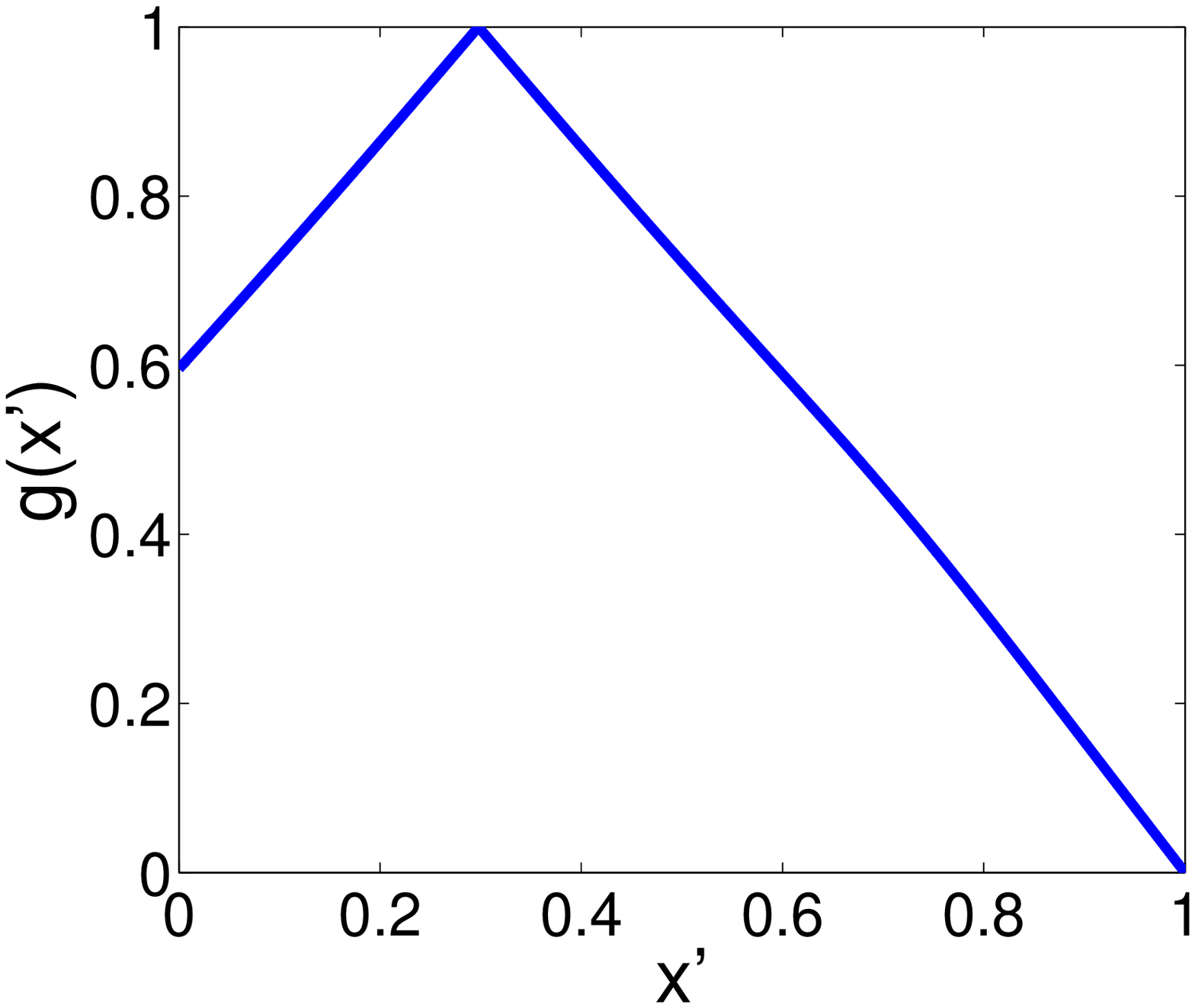}}
\caption{The graph of (a) the map with three measure singularities and (b) its conjugate map. }
\label{fig:21}
\end{figure}

\begin{figure}[htp]
\subfigure[the natural measure of the map with measure three singularities]{\includegraphics[width=0.48\textwidth,height=0.3\textwidth]{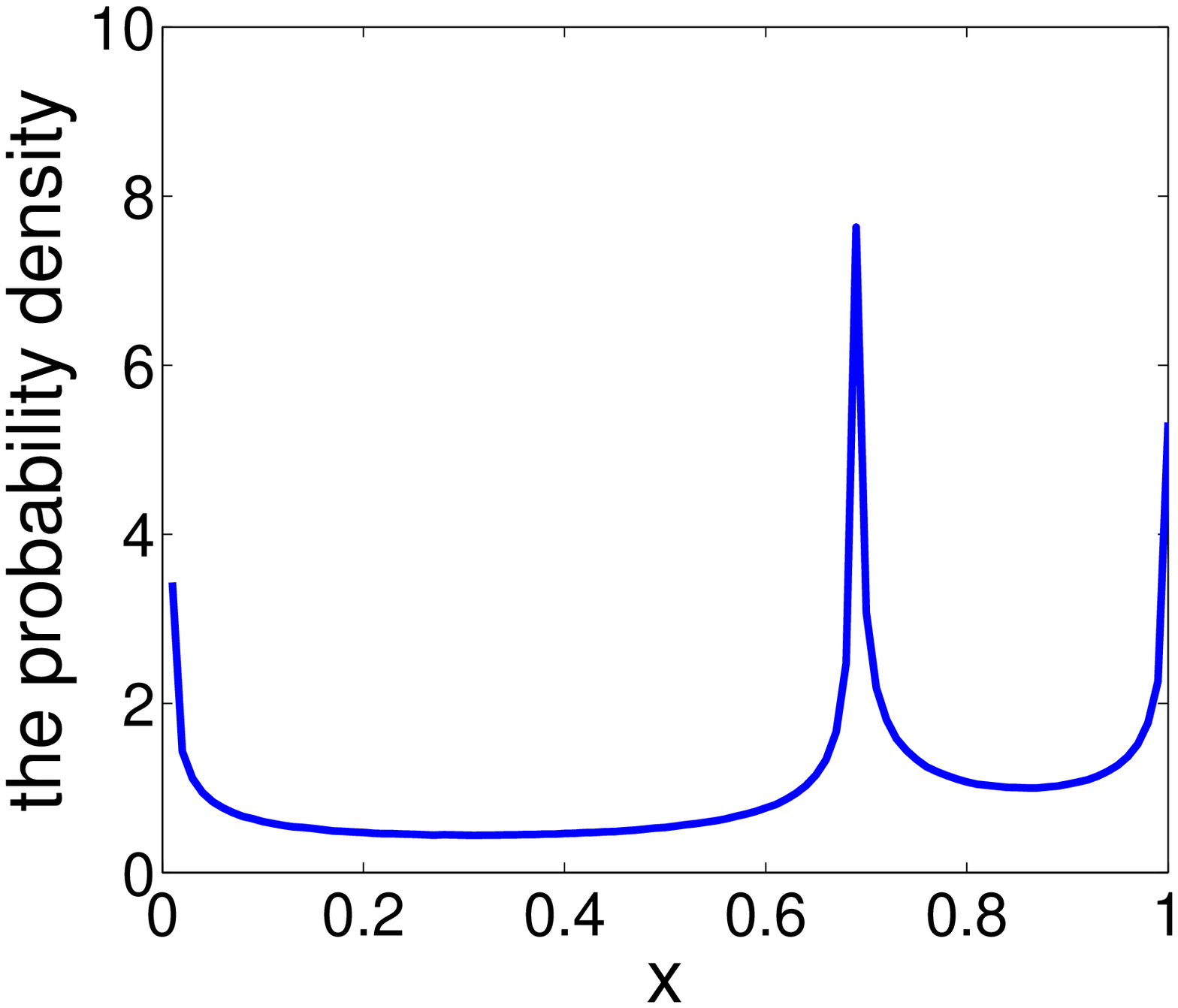}}
\subfigure[the natural measure of the conjugate map]{\includegraphics[width=0.48\textwidth,height=0.3\textwidth]{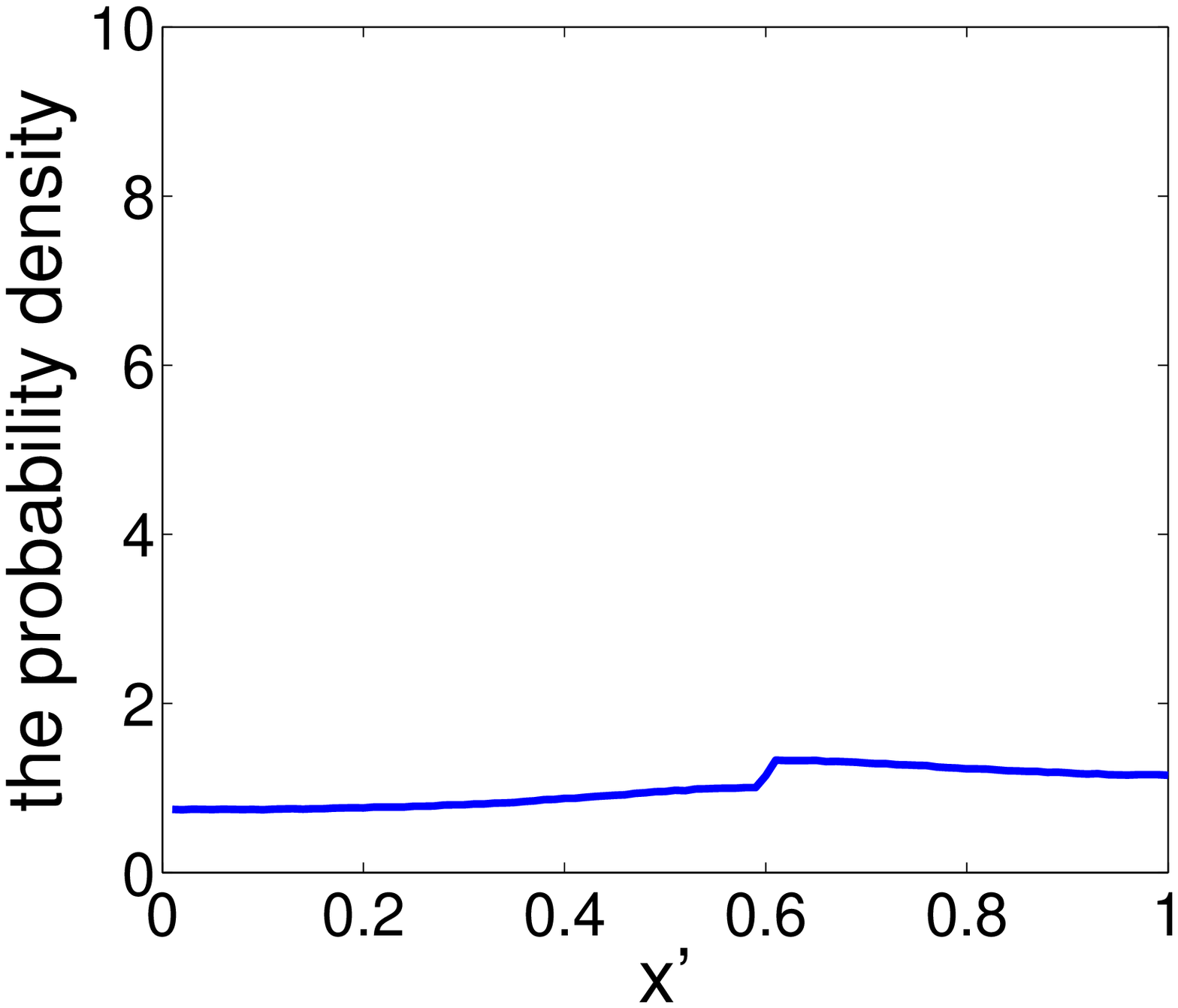}}
\caption{The natural measure of (a) the map with three measure singularities and (b) its conjugate map.}
\label{fig:22}
\end{figure}

\begin{figure}[htp]
\includegraphics[width=0.48\textwidth,height=0.3\textwidth]{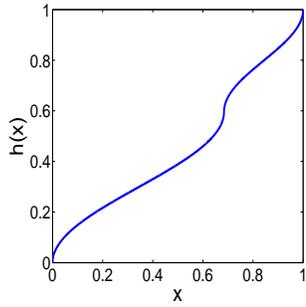}
\caption{The conjugacy $h(x)$ for the map with three measure singularities.}
\label{fig:23}
\end{figure}

The maps we have studied so far all have two measure singularities: $x=0$ and $x=1$. In this section, we turn to a map with three measure singularities, the graph of which is shown in FIG.~\ref{fig:21}(a). The exact functional form of the map is $f(x)=\sin(\frac{\pi}{a}(1-x))$, where $a=1.3156445888...$. It has a critical point of order two, and has a nice property: $f(0)=x_f$, where $x_f$ is the unique fixed point of the map. The natural measure of $f$ has three singularities: $x=0,x_f,1$. The asymptotic form of the natural measure near the singularity is: $\rho \sim \frac{1}{\sqrt{x}}$ near $x=0$, $\rho \sim \frac{1}{\sqrt{|x-xf|}} $ near $x=x_f$ and $\rho \sim \frac{1}{\sqrt{1-x}}$ near $x=1$, as shown in FIG.~\ref{fig:22}(a). To clear out the singularities, we use the coordinate transformation $h(x)$ as depicted in FIG.~\ref{fig:23}, which stretches the coordinate around the singularities. The conjugate map $ g(x')$ is nearly a piecewise linear map, as shown in FIG.~\ref{fig:21}(b). The natural measure of the  map $g(x')$ is depicted in FIG.~\ref{fig:22}(b), which has no singularity any more.


\begin{figure}[htp]
\subfigure[the  error of escape rate]{\includegraphics[width=0.48\textwidth,height=0.3\textwidth]{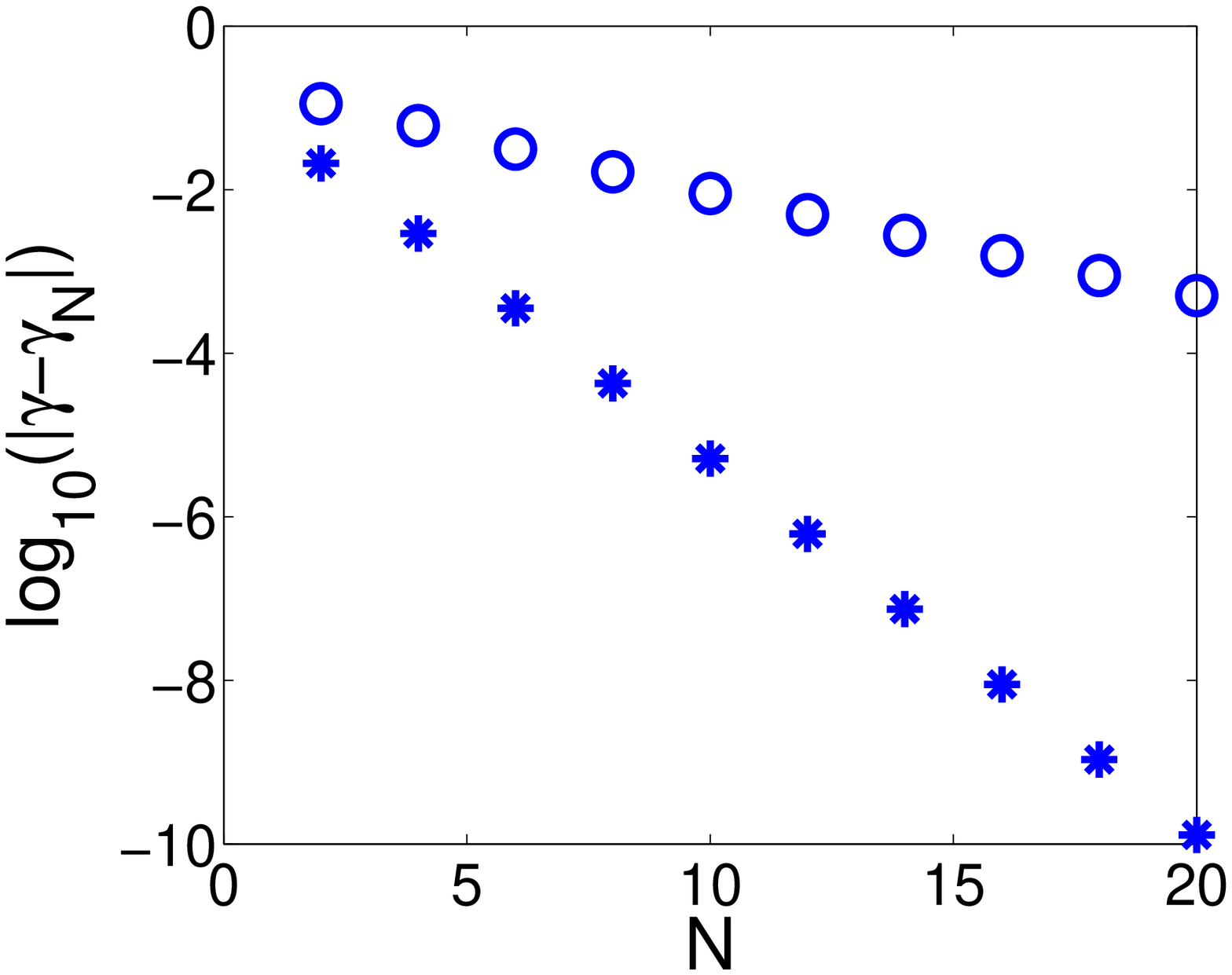}}
\subfigure[the  error of $\langle x \rangle$]{\includegraphics[width=0.48\textwidth,height=0.3\textwidth]{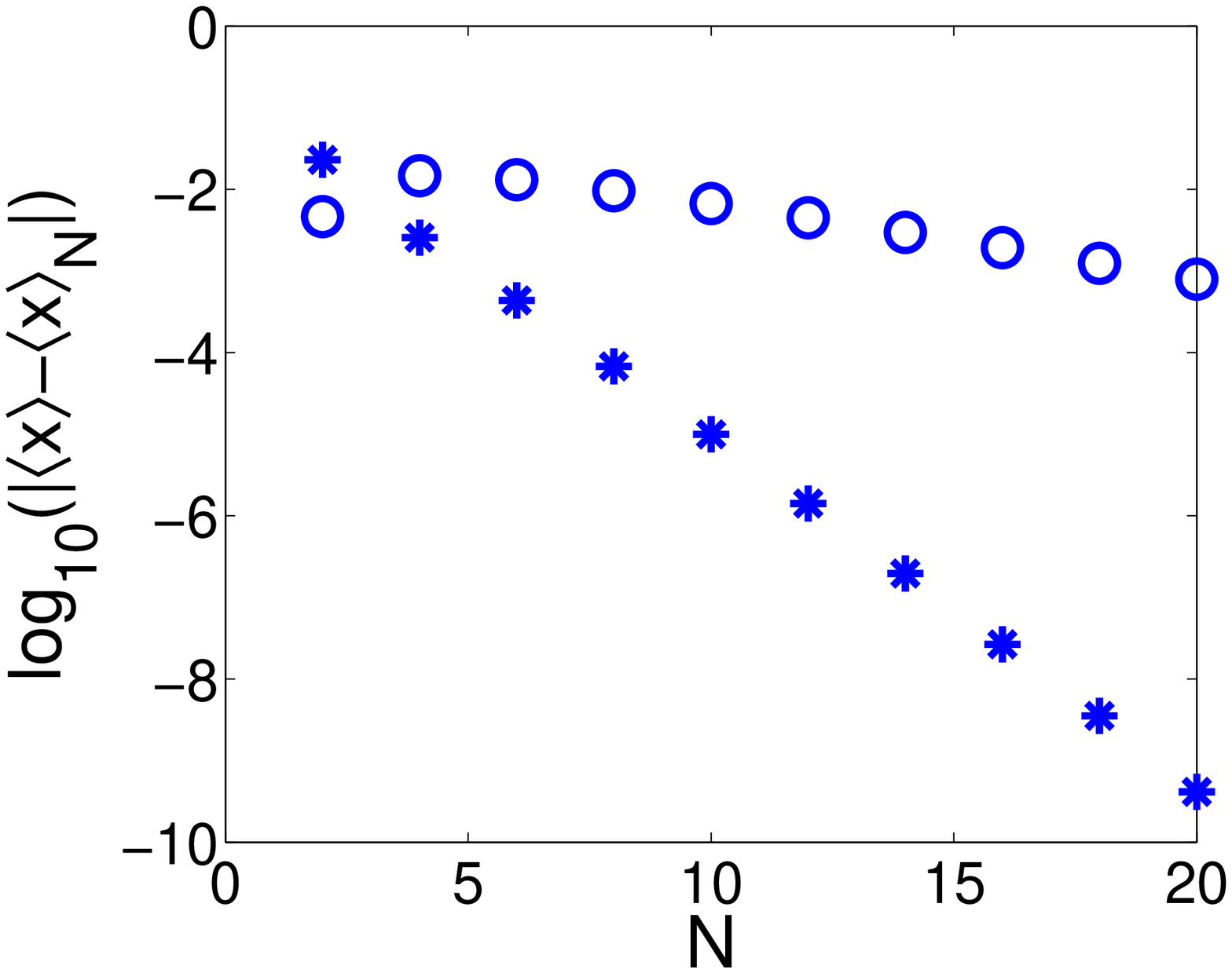}}
\subfigure[the  error of $\langle x^2 \rangle$]{\includegraphics[width=0.48\textwidth,height=0.3\textwidth]{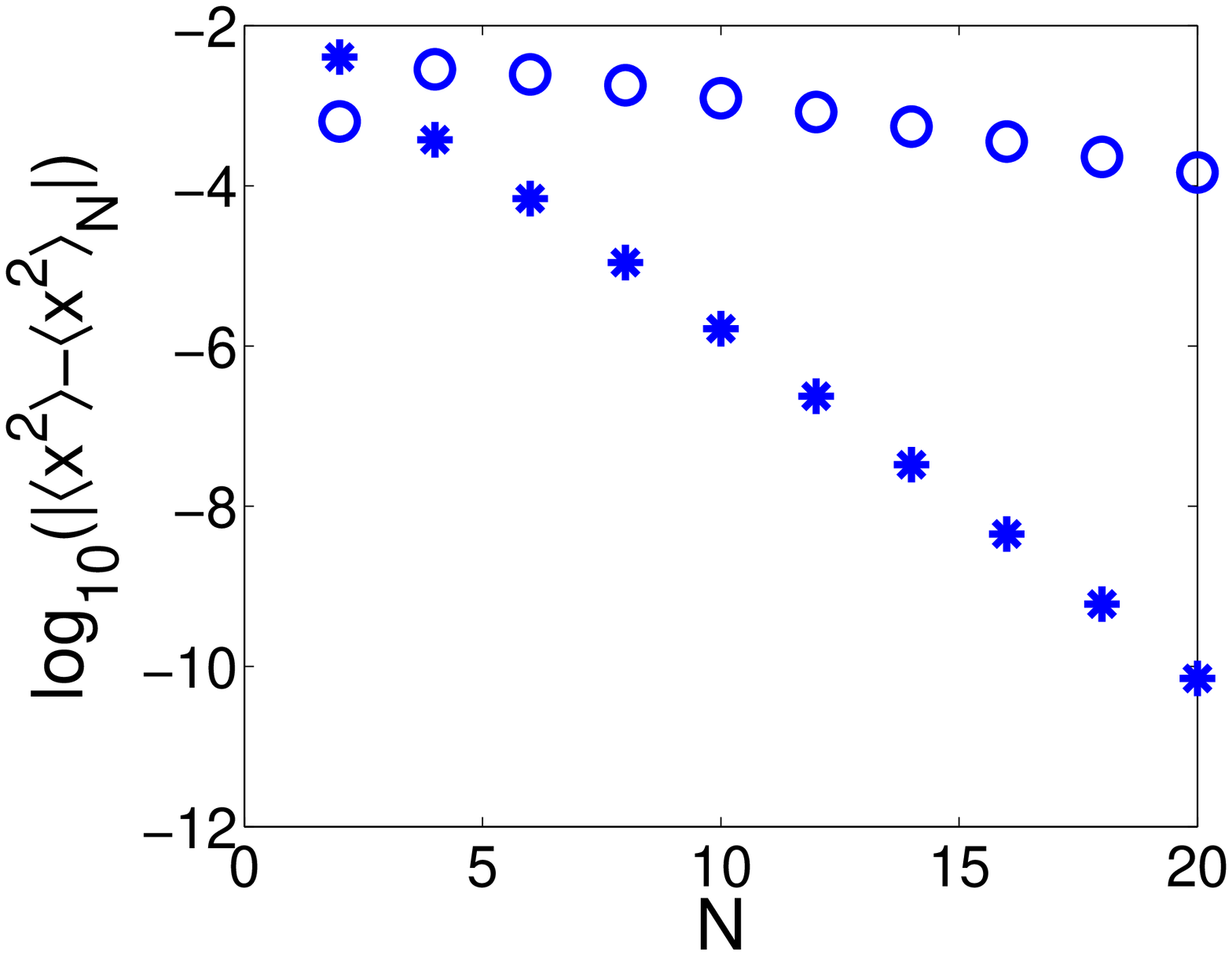}}
\subfigure[the  error of $\langle x^3 \rangle$]{\includegraphics[width=0.48\textwidth,height=0.3\textwidth]{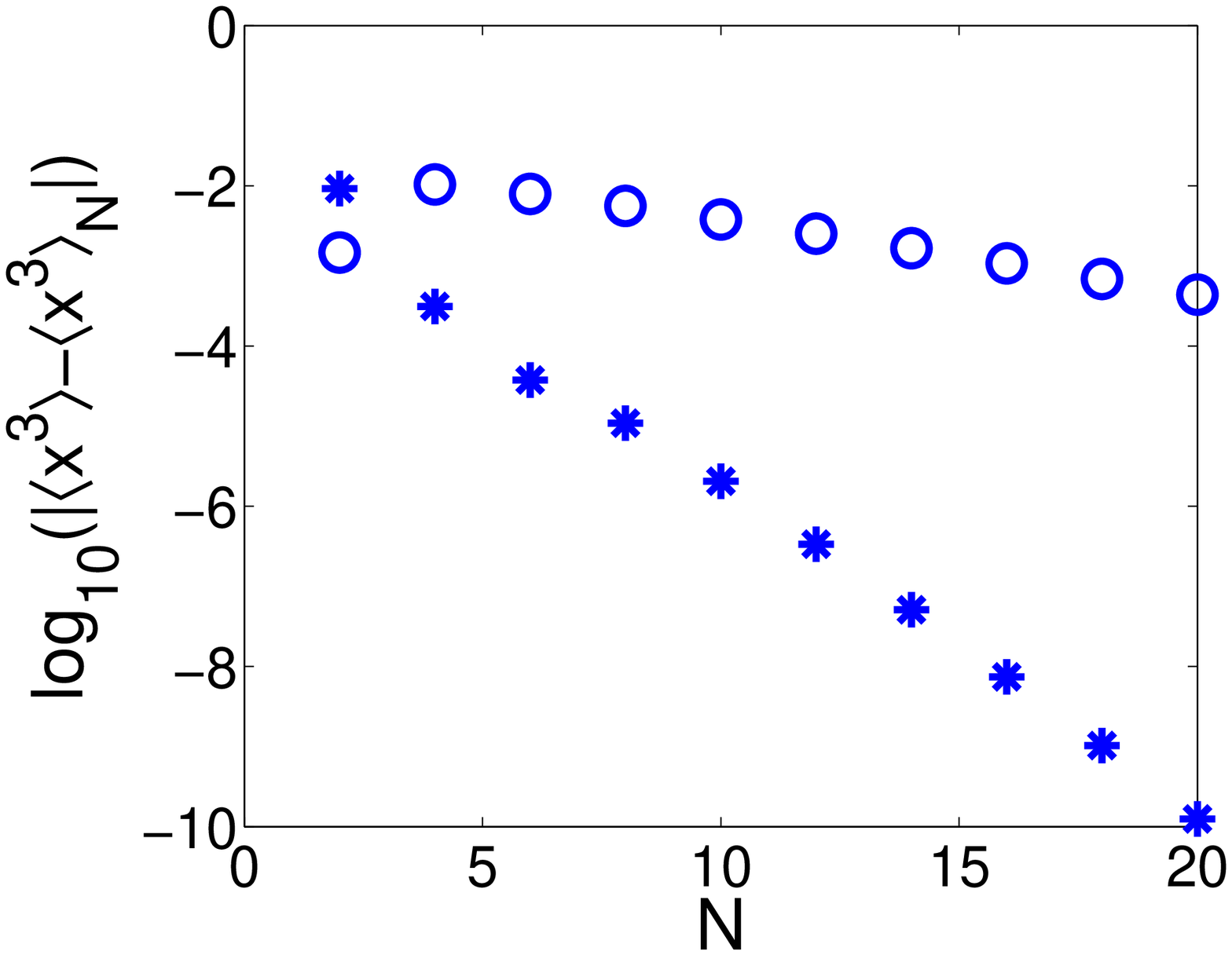}}
\caption{The  error of the escape rate, $\langle x \rangle\,,\langle x^2 \rangle\,,\langle x^3\rangle$ obtained by the dynamical zeta function(circles) for the map with three measure singularities and its conjugate dynamical zeta function(stars).}
\label{fig:24}
\end{figure}

For the conjugate dynamical zeta function, the stability of $\overline{x_f}$ should be changed to $|\Lambda_{x_f}|'=|\Lambda_{x_f}|^{\frac{1}{2}}$. Again, we use the original and the conjugate dynamical zeta function to calculate averages. The results  are listed in TABLE~\ref{ta:6}, with a cutoff of cycle length $20$.  The errors in the computation are plotted in FIG.~\ref{fig:24}. Note that the binary symbolic dynamics is not complete in the current example. The number of cycles get much reduced compared with the full symbolic dynamics case.

By clearing out the singularities in the natural measure, the convergence is accelerated a lot. So, in this case, the conjugate dynamical zeta function is still an effective way to acquire averages with high accuracy.
\begin{table}[htp]
\begin{tabular}{|c|c|c|}
\hline
 & the dynamical& the conjutate\\
 &  zeta function& dynamical zeta function\\
 \hline
 escape rate & $5\times 10^{-4}$ & $-1\times10^{-10}$\\
 \hline
 $\langle x \rangle$ & $0.601$ & $0.601895610$ \\
  \hline
  $\langle x^2 \rangle$ & $ 0.453$ & $0.453165976$\\
  \hline
  $\langle x^3 \rangle$ & $0.36$ & $0.364669939 $\\
  \hline
\end{tabular}
\caption{The averages for the map with three measure singularities computed with two different methods.}
\label{ta:6}
\end{table}

\subsubsection{A map with measure singularities on a period-2 orbit}
\begin{figure}[htp]
\subfigure[the map with measure singularities on a period-2 orbit]{\includegraphics[width=0.48\textwidth,height=0.3\textwidth]{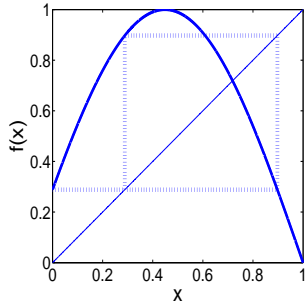}}
\subfigure[the conjugate map]{\includegraphics[width=0.48\textwidth,height=0.3\textwidth]{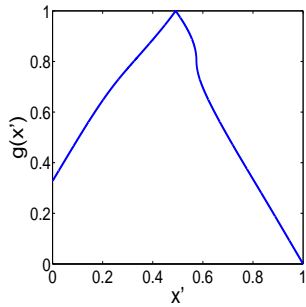}}
\caption{The graph of (a) the map with measure singularities on a period-2 orbit and (b) its conjugate map.}
\label{fig:25}
\end{figure}

\begin{figure}[htp]
\subfigure[the natural measure of map $f$]{\includegraphics[width=0.48\textwidth,height=0.3\textwidth]{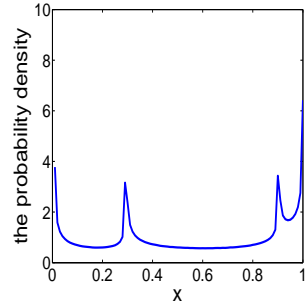}}
\subfigure[the natural measure of the conjugate map]{\includegraphics[width=0.48\textwidth,height=0.3\textwidth]{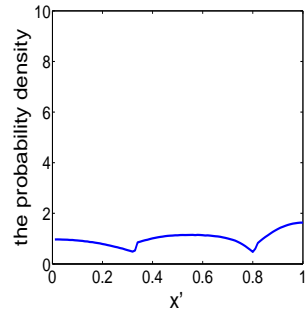}}
\caption{The natural measure of (a) the map with measure singularities on a period-2 orbit  and (b) its conjugate map.}
\label{fig:26}
\end{figure}

\begin{figure}[htp]
\includegraphics[width=0.48\textwidth,height=0.3\textwidth]{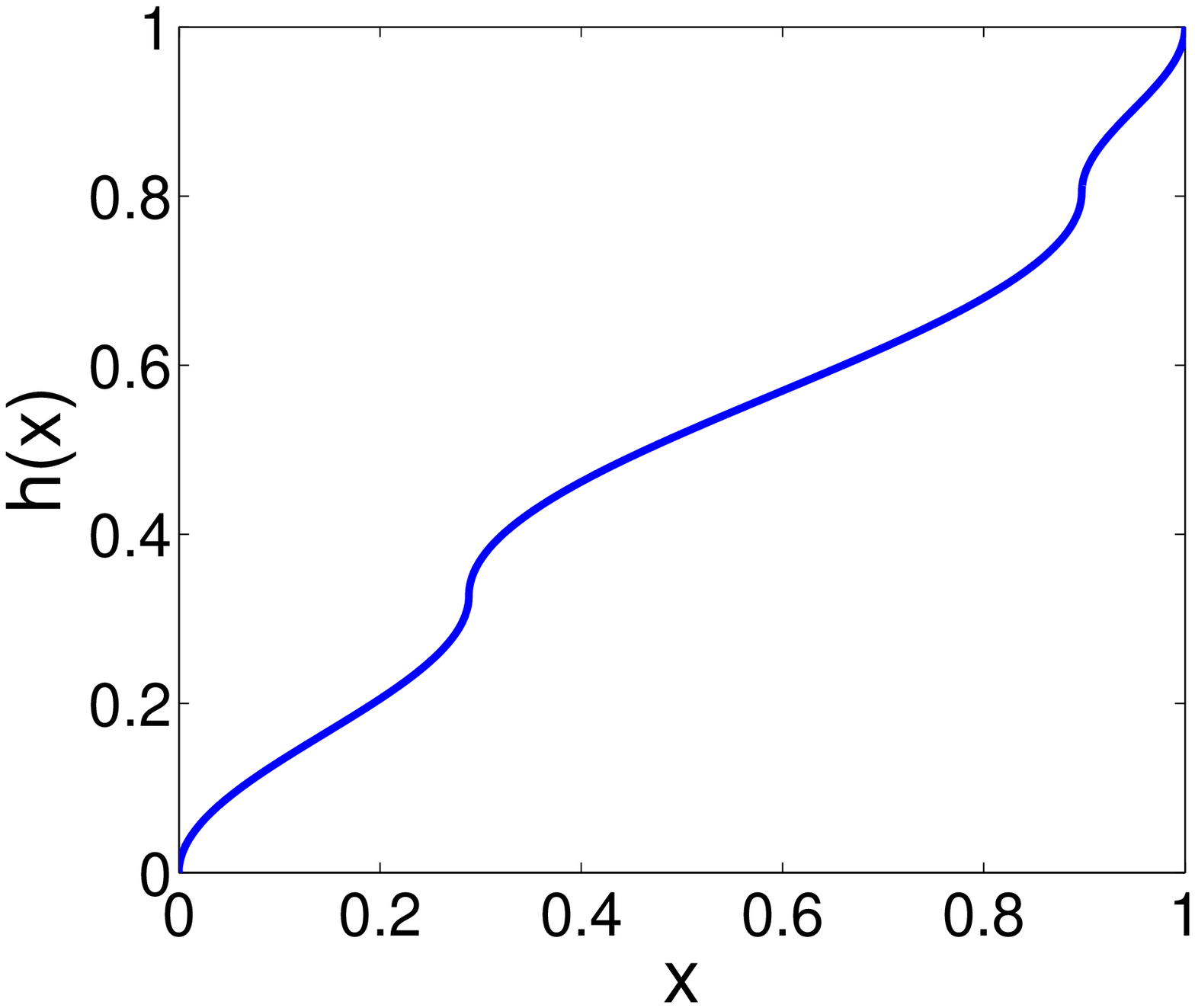}
\caption{The conjugacy $h(x)$ for the map with measure singularities on a period-2 orbit.}
\label{fig:27}
\end{figure}

 The conjugation method may be applied to systems with measure singularity on longer orbits. As an example, we study a map with measure singularities on a period-2 orbit. The functional form of the map is still $f(x)=\sin(\frac{\pi}{a}(1-x)) $, but with a different value $a=1.10263451544766... $. FIG.~\ref{fig:25}(a) portrays the graph of the map. For this map, $ f(0)=x_a,\,f^2(0)=x_b$, where $x_a$ and $x_b$ make a period-2 orbit. The natural measure of the map is shown in FIG.~\ref{fig:26}(a) and has  four singularities: $x=0,\,x_a,\,x_b,\,1$. An appropriate conjugacy $h(x)$ is plotted in FIG.~\ref{fig:27}, which stretches the coordinate around the four singularities. The conjugate map $g(x')$ and its natural measure are plotted in FIG.~\ref{fig:25}(b) and FIG.~\ref{fig:26}(b) respectively. For the map $g(x')$, the singularities has been removed---the situation that we have experienced many times.

\begin{figure}[htp]
\includegraphics[width=0.48\textwidth,height=0.3\textwidth]{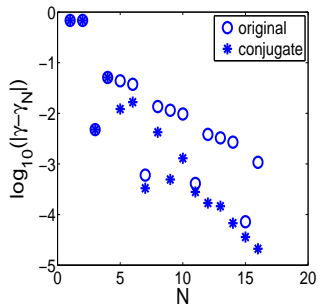}
\caption{The error of the escape rate by the original(circles) and the conjugate(stars) dynamical zeta function for the map with measure singularities on a period-2 orbit.}
\label{fig:28}
\end{figure}

We change the eigenvalue of the period-2 orbit and obtain the conjugate dynamical zeta function---just as what we have done.  The conjugate dynamical zeta function gives interesting results in the calculation of the escape rate.  FIG.~\ref{fig:28} shows the error in the escape rate  computed with the original and the conjugate dynamical zeta function. In general, the results obtained with the conjugate dynamical zeta function converge exponentially and uniformly, faster than with the original dynamical zeta function. However, it's not totally true here.  The conjugate dynamical zeta function doesn't accelerate the convergence as effectively as before.  So, why the convergence for the conjugate dynamical zeta function is not so good even if we have removed the measure singularities?

\begin{figure}[htp]
\includegraphics[width=0.48\textwidth,height=0.3\textwidth]{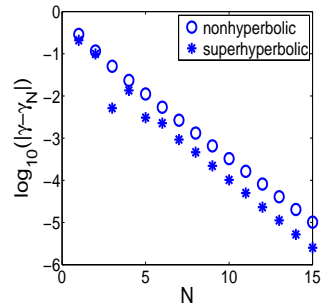}
\caption{The  error of the escape rate for the super-hyperbolic map and the logistic map.}
\label{fig:29}
\end{figure}

If we stare at FIG.~\ref{fig:25}(b), the graph of the conjugate map, we find one special point at which the slope is infinite. It's just this point that causes the slow convergence. We regard  this infinite slope map as ``super-hyperbolic''. The  super-hyperbolic map can slow down the convergence. To illustrate this point, we use the dynamical zeta function to calculate the escape rate of a super-hyperbolic map $f(x)$,
\begin{equation}
f(x)= \{ \begin{array}{ll} 2x & x \in [0,1/2] \\ \sqrt{2-2x} & x \in [1/2,1]\,,\end{array}
\end{equation}
where $f'(1)=\infty$. We compare the results obtained from this map and the logistic map in FIG.~\ref{fig:29}. Both results are computed with the original dynamical zeta function. We can see that the convergence rates are similar for these two maps. So, the super-hyperbolicity is harmful to the convergence, just like the non-hyperbolicity. To understand this, we recall the fact that the average obtained by the truncated dynamical zeta function is nearly identical to the one computed with the corresponding piecewise linear map. So, if the piecewise linear map can approximate the original map very well, the average obtained would be quite accurate. However, for the super-hyperbolic map, there exists a point with infinite slope, which means that the value of the map changes extremely unevenly near that point. So, a lot of cycle points are needed near that point to get a fair approximation just like near the critical of a non-hyperbolic map. Thus, the slow convergence of the super-hyperbolic map is expected.

Based on the discussion above, we can see that the singularity in the natural measure is not the only factor which  influences the convergence of the dynamical zeta function. For the map with measure singularities on a period-2 orbit, clearing out the measure singularities doesn't improve the convergence much. In fact, the  direct and  essential factor to determine the convergence of the dynamical zeta function is the cancelation between prime cycles and pseudo-cycles. For the super-hyperbolic case, the cancelation is not good even if the measure singularities do not exist. So, in the super-hyperbolic case,  how to incur further cancelation  remains a challenging problem.

\section{Conclusion \label{sec:sum}}

 The central idea of this paper is that by clearing
out the singularities in the natural measure we may accelerate the convergence of cycle expansions. Maps with critical points produce natural measures with singularities and show bad convergence in the expansion calculation. With appropriate coordinate transformation, the resulted conjugate map produces no singularity in its natural measure. To calculate dynamical averages, we use the conjugate spectral function, the cycle expansion of which is greatly accelerated due to the removal of the singularities. Essentially, the method locates leading poles of the spectral function by classifying singularities in the natural measure and removes them through a coordinate transformation.

We test our method on several maps, {\em i.e.},
$f(x)=1-|2x-1|^k,\,k=2,4,6$, $f(x)=\sin(\pi x)$ which have only one critical point and complete binary dynamics, and the map which has three measure singularities. For these maps, the conjugate dynamical zeta function converges much faster than the original zeta function. Also, the conjugate spectral determinant restores the super-exponential convergence. However, when we treat the map with measure singularities on a period-two orbit, we find that the conjugate dynamical zeta function does not converge as fast as
expected. Analysis shows that the super-hyperbolicity of the
conjugate map leads to this slowing-down. Further study is needed to eliminate this nuisance.

In this paper, we use one-dimensional maps as examples to
demonstrate our accelerating scheme. How to generalize it to higher dimensions or to flows requires further investigation. Even in the 1-d case, we can can only treat maps with symbolic dynamics being subshifts of finite type. If the genealogy sequence of the critical point is not essentially periodic, there may exist a natural boundary in the complex plane for the dynamical zeta function on which singular points are dense. In this case, the natural measure is singular on a dense and countable set~\cite{dah97prun}. It seems not possible to expand the radius of expansion by analytic continuation. Novel techniques have to be invented to achieve accelerated convergence in this case.

\section*{Acknowledgements}
This research is supported by National Natural Science Foundation of
China (Grant No. 10975081) and the ph.D. Programs Foundation of
Ministry of Education of China (Grant No.20090002120054).

\bibliography{nonlind}
\bibliographystyle{plain}

\end{document}